\documentclass[pre,amsmath,onecolumn, showpacs, superscriptaddress,10pt]{revtex4-1}
\usepackage{amssymb}
\usepackage{amsmath}
\usepackage{hyperref}
\usepackage{graphicx}

\usepackage{color}
\definecolor{Blue}{rgb}{0.00, 0.00, 1.00}
\definecolor{Red}{rgb}{1.00, 0.00, 0.00}

\newcommand{\bea}{\begin{eqnarray}}
\newcommand{\eea}{\end{eqnarray}}

\begin{document}

\title{Statistics of the maximal distance and momentum in a trapped Fermi gas at low temperature}

\author{David S. \surname{Dean}}
\affiliation{Univ. Bordeaux and CNRS, Laboratoire Ondes et Mati\`ere  d'Aquitaine
(LOMA), UMR 5798, F-33400 Talence, France}
\author{Pierre Le Doussal}
\affiliation{CNRS-Laboratoire de Physique Th\'eorique de l'Ecole Normale Sup\'erieure, 24 rue Lhomond, 75231 Paris Cedex, France}
\author{Satya N. \surname{Majumdar}}
\affiliation{Univ. Paris-Sud, CNRS, LPTMS, UMR 8626, Orsay F-91405, France}
\author{Gr\'egory \surname{Schehr}}
\affiliation{Univ. Paris-Sud, CNRS, LPTMS, UMR 8626, Orsay F-91405, France}

\begin{abstract} 
We consider $N$ non-interacting fermions in an isotropic $d$-dimensional harmonic trap. We compute analytically
the cumulative distribution of the maximal radial distance of the fermions from the trap center at zero temperature. 
While in $d=1$ the limiting distribution (in the large $N$ limit), properly centered and scaled, converges to the squared
Tracy-Widom distribution of the Gaussian Unitary Ensemble in Random Matrix Theory, we show
that for all $d>1$, the limiting distribution converges to the Gumbel law. These limiting forms turn out
to be universal, i.e., independent of the details of the trapping potential for a large class of isotropic trapping potentials. We also study the position of the right-most fermion in a given direction in $d$ dimensions and, in the case of a harmonic trap, the maximum momentum, and show that they obey similar Gumbel statistics. Finally, we generalize these results to low but finite temperature.
%The statistics of the distance of the fermion farthest from the origin is analysed.  
%In contrast to the case of one dimension the statistics of the maximal fermion distance from the origin
%is not described by  the Tracy Widom distribution but by  a deterministically translated  Gumbel distribution. In higher dimensions the stationary states of the Schr\"odinger equation can be characterized by their total angular momentum $l$,
%which is integer valued. The maximum distance in each sector of angular momentum $l$ is independent of 
%the others, however the distribution for each sector $l$ depends explicitly on $l$ and so although the maximal
%fermonic distance in each sector is independent of the others they do not have the same distribution.
\end{abstract}


\maketitle
\section{Introduction}

Recent experimental advances in cold atom systems, in one or higher dimensions, have raised interesting new
questions about these basic and fundamental systems~\cite{BDZ08,GPS08}. One such simple system corresponds to $N$ non-interacting spin-less fermions in a $d$-dimensional harmonic trap. Near the center of the trap the system is just a free Fermi gas, which is a very well studied system in condensed matter physics and it can be analyzed using standard methods such as the Local Density Approximation (LDA) \cite{Mahan,Castin}. However, the trap confines the Fermi gas in a limited region of space and thus
introduces a sharp edge in the Fermi gas, in the limit of large $N$. Near this edge, the density is small and consequently quantum and thermal fluctuations are strong, leading to new emergent edge physics. The nature of this physics at the edges  has been the subject of a number of  recent studies \cite{Kohn, eis13,marino_prl,dea15a, dea15b,marino_pre,dea16}.

The system  of interest here  consists of $N$ non-interacting spin-less fermions in an isotropic $d$-dimensional harmonic potential $V({\bf x}_1, \cdots, {\bf x}_N) = \frac{1}{2}m\omega^2 \sum_{i=1}^N {\bf x}_i^2 $, where ${\bf x}_i$ denotes the position of the $i$-th fermion in $d$ dimensions. In the following, for convenience, we will set $m=\omega = \hbar = 1$. The main  focus of this paper will be on quantum fluctuations, i.e., setting the temperature $T=0$, however some preliminary results on the role of finite temperature will also be given. At $T=0$, the system will be in its many-body quantum ground state with the ground-state wave function denoted by $\Psi_0({\bf x}_1, {\bf x}_2, \cdots, {\bf x}_N)$. The quantum fluctuations are then characterized by the joint probability distribution function (JPDF), $P({\bf x}_1, {\bf x}_2, \cdots, {\bf x}_N) = |\Psi_0({\bf x}_1, {\bf x}_2, \cdots, {\bf x}_N)|^2$, normalized to unity. 

In $d=1$, this many-body ground state wave function can be computed explicitly \cite{eis13,marino_prl,meh91}
\bea
P({x}_1, {x}_2, \cdots, {x}_N) = |\Psi_0({x}_1, {x}_2, \cdots, {x}_N)|^2 = \frac{1}{Z_N} \prod_{1\leq i < j \leq N}(x_i-x_j)^2 \, e^{-\sum_{i=1}^N x_i^2} \label{1d_GUE} \;,
\eea
where $Z_N$ is a normalization constant. Thus in $1d$ the positions $x_i$'s behave exactly as the eigenvalues of a random matrix belonging to the Gaussian Unitary Ensemble (GUE). In this case, the average density of fermions (normalized to unity) defined as $\rho_N(x) = (1/N)\sum_{i=1}^N \langle \delta(x-x_i)\rangle_0$ (where $\langle \ldots \rangle_0$ refers to a ground state average with respect to the JPDF in Eq. (\ref{1d_GUE})) converges in the large $N$ limit to the celebrated Wigner semicircular form \cite{meh91,for10,fyodorov}
\bea\label{Wigner_intro}
\rho_N(x) = \frac{1}{\sqrt{N}} f_W\left(\frac{x}{\sqrt{N}}\right) \;\;, \;\; f_W({\sf x}) = \frac{1}{\pi}\sqrt{2-{\sf x}^2} \;.
\eea 
Thus, on an average, the fermions are confined in a finite interval of the real line, $[-x_{\rm edge}, +x_{\rm edge}]$ where $x_{\rm edge} = \sqrt{2\,N}$. Another striking prediction of this mapping \cite{dea15a} is that the quantum fluctuations of the position of the rightmost fermion at $T=0$, i.e., the probability distribution of $x_{\max} = \max\{x_1, x_2, \cdots, x_N \}$, converges in the large $N$ limit to 
the Tracy-Widom (TW) distribution characterizing the distribution of the largest eigenvalue of a GUE random matrix \cite{TW94}.

In $d>1$, while the direct connection to Random Matrix Theory (RMT) no longer holds, it has been shown recently that several methods from RMT can still be exploited to study the correlations of the Fermi gas in $d>1$ \cite{dea15b,dea16}. Notably, in the large $N$ limit, there is a determinantal structure in the problem which allows one to express any ground-state correlation function as a determinant involving a kernel, for all $d \geq 1$. This kernel has been analyzed in detail in the limit of large $N$ \cite{dea16}. For instance, the average density of the fermions, which is spherically symmetric, converges in the large $N$ limit  to
\bea\label{density_d_intro}
\rho_N({\bf x}) = \rho_N(r) \simeq \frac{1}{N (2\sqrt{\pi})^d \Gamma(\frac{d}{2}+1)}\, \left(r_{\rm edge}^2 - r^2\right)^{d/2} \theta(r_{\rm edge} - r) \:,
\eea  
with
\bea\label{redge_intro}
r_{\rm edge} = 2^{1/2} \,\left[\Gamma(d+1) \right]^{\frac{1}{2d}} N^{\frac{1}{2d}} = \sqrt{2\,\mu} \;,
\eea
where $\mu$ denotes the Fermi energy. The average density in Eq. (\ref{density_d_intro}) is a generalization of the Wigner semicircular law in $d \geq 1$. Thus, like in $d=1$, the Fermi gas is confined in a finite sphere of radius $r_{\rm edge}$ around the trap center. As mentioned earlier, in $d=1$, the observable $x_{\max}$ (the position of the rightmost fermion) obeys the Tracy-Widom distribution in the large $N$ limit. It is then natural to ask what is the analogue of this observable in higher dimensions and what is its limiting distribution in the large $N$ limit?

\begin{figure}[ht]
\includegraphics[width = 0.4\linewidth]{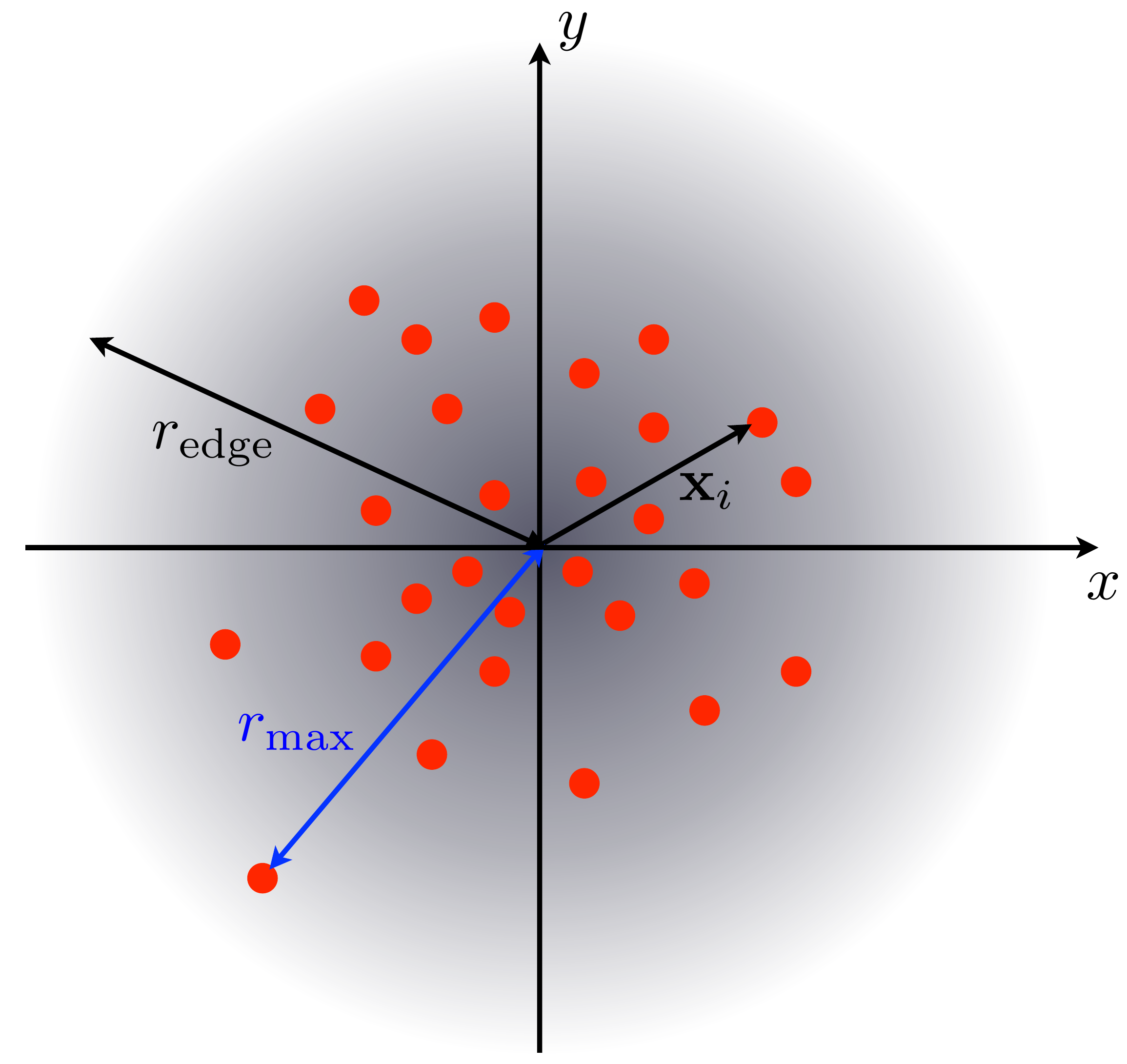}
\caption{Sketch of one configuration of $N$ non-interacting fermions in a two-dimensional isotropic harmonic trap, centered at the origin, at zero temperature. The shaded area indicates the average density profile $\rho_N(r)$ (\ref{density_d_intro}), which vanishes beyond $r_{\rm edge}$ in the large $N$ limit. The fermion positions are labeled by ${\bf x}_i$, with $i=1, \cdots, N$ and their JPDF is given by $P({\bf x}_1, {\bf x}_2, \cdots, {\bf x}_N) = |\Psi_0({\bf x}_1, {\bf x}_2, \cdots, {\bf x}_N)|^2$, with $\Psi_0({\bf x}_1, {\bf x}_2, \cdots, {\bf x}_N)$ the many-body ground state wave function. Here we focus on the fluctuations of the radial distance $r_{\max}$ of the farthest fermion (\ref{rmax_intro}). In the large $N$ limit, we show here that the PDF of $r_{\max}$, properly centered and scaled converges to a Gumbel distribution [see Eq. (\ref{result_1})].}\label{Fig_rmax}
\end{figure}
A natural observable in higher dimensions, that generalizes the position of the rightmost fermion for $d=1$, is 
the radial distance of the farthest fermion    
\begin{eqnarray}\label{rmax_intro}
r_{\max} = \max(r_1, r_2, \ldots, r_N) \;\;\; {\rm where} \;\;\; r_i^2 = {{\bf x_i}}\cdot {\bf x_i} \;.
\end{eqnarray}
In the ground state, with JPDF $P({\bf x}_1, {\bf x}_2, \cdots, {\bf x}_N)$, $r_{\max}$ is clearly a random variable and in this paper our main focus will be to study the cumulative probability  
\begin{eqnarray}\label{cdf_intro}
P(w,N) = {\rm Prob.}(r_{\max} \leq w,N)  \;,
\end{eqnarray} 
 in all dimensions. Although we will obtain a formal solution for arbitrary $N$, the most
interesting question concerns the large $N$ limit.
%in the limit of large $N$ and in all dimensions. 
The answer to this question is straightforward in $d=1$ where, using the exact mapping to RMT mentioned above, it is easy to show that for large $N$ the limiting form of $P(w,N)$ converges to the {\it squared} TW distribution (see the discussion later). What is the analogous limiting form in $d>1$? The main purpose of this paper is to address this question and present an exact result for this limiting distribution for all $d>1$. 

Let us summarize our main results. In the limit of large $N$ and $d>1$, we show that this cumulative distribution converges to the limiting scaling form
\bea\label{result_1}
P(w,N) \underset{N \to \infty}{\longrightarrow} G\left(\frac{w-A_N}{B_N} \right) \;, \; {\rm where} \;\; G(z) = e^{-e^{-z}} \;,
\eea
where the scale factors $A_N$ and $B_N$ can be computed explicitly for large $N$. To leading order for large $N$, we show that
\bea\label{AN_BN}
A_N \simeq r_{\rm edge} + \frac{1}{\sqrt{2}\mu^{1/6}} a_\mu \;\;\quad {\rm and} \;\;\quad B_N \simeq \frac{1}{2 \sqrt{2 \, a_\mu \, }\mu^{1/6}} \; ,
\eea
where $r_{\rm edge} = \sqrt{2 \mu}$ denotes the edge of the Fermi gas, $\mu \simeq \left( \Gamma(1+d)\,N\right)^{1/d}$ is the Fermi energy and $a_\mu$ is given by
\bea
\label{result_2}
&&a_\mu \simeq  \left(\frac{d-1}{2} \ln{\mu}\right)^{2/3} \;.
\eea
The limiting distribution $G(z)$ in Eq. (\ref{result_1}) is a Gumbel distribution.  In addition, we
also study the distribution of $x_{\max}$, the position of the rightmost fermion in $d$ dimensions,
and again find a limiting Gumbel law. We also investigate the effect of finite temperature and
show again the emergence of a Gumbel distribution for these observables ($r_{\max}$ and $x_{\max}$)
although with different parameters.

It is well known that the Gumbel distribution emerges as a limiting distribution in the classical theory of extreme value statistics of independent and identically distributed (i.i.d.) random variables \cite{gum58,gal87}. In contrast, in the present case of fermions in a $d$-dimensional confining potential, the positions of the fermions are strongly correlated due to the Pauli exclusion principle. Thus it appears as a puzzle as to why the same Gumbel distribution for i.i.d. variables appears in this case. By decomposing the wave function into radial and angular coordinates, we will indeed unveil a ``decorrelation'' mechanism which explains the emergence of the Gumbel distribution, even in this strongly correlated system. It is interesting to note that a similar Gumbel limiting form also appears in the context of Ginibre matrices \cite{meh91,for10,gin65,KS11}, which are matrices with i.i.d. Gaussian entries (real, complex or quaternionic), but without any special symmetry of the matrix. In this case, the eigenvalues are generally distributed over the complex plane and the distribution of the farthest eigenvalue from the origin, properly centered and scaled, converges for large $N$ to a Gumbel distribution \cite{rid14,rid03} (see also \cite{CMV2016} for a study of the large deviations beyond the Gumbel). For complex Ginibre matrices, the universality of the Gumbel law (with respect to different confining potentials), was established in Ref. \cite{CP2015}. The ``decorrelating'' mechanism for the Ginibre matrices thus appears to be similar to that for the fermion problem studied here.

One interesting property of the harmonic oscillator is that the position ${\bf x}$ and momentum ${\bf p}$ play completely symmetric roles. Hence, although we will not develop it here, all of our results for $r_{\max}$ also carry out to $p_{\max}$, the momentum of the fermion with the maximal momentum 
\bea
p_{\max} = \max(p_1, p_2, \ldots, p_N) \;\;\; {\rm where} \;\;\; p_i^2 = {{\bf p_i}}\cdot {\bf p_i} \;.
\eea 
Hence we predict that this quantity will be governed by a Gumbel distribution in $d>1$ for large $N$
(while it involves the Tracy-Widom distribution for $d=1$).

The paper is organized as follows. In section II, we recapitulate the general determinantal properties of non-interacting fermions in a harmonic potential and present the first derivation of the limiting distribution of the radial distance of the farthest fermion in $d>1$, using the asymptotic properties of an underlying Fredholm determinant. In section III, we decompose the ground state wave function into radial and angular coordinates and present a second derivation of the same limiting distribution, that demonstrates the emergence of a ``decorrelation'' mechanism. In section IV, we discuss  some preliminary extensions of our results and, in section V, we give our conclusions about this study.

\section{$N$ free fermions in a $d$-dimensional harmonic potential: ground state}

\label{sec:GS}

\subsection{Determinantal process}

We consider $N$ spin-less free fermions in a $d$-dimensional harmonic potential. The many-body Hamiltonian
is given~by
\begin{eqnarray}\label{def_H}
\hat H = \sum_{i=1}^N \hat h_i \;, \; \;\;\; {\rm where} \;\;\;\; \hat h_i = -\frac{1}{2}\nabla_{{\bf x}_i}^2 + \frac{1}{2} r_i^2 \;, 
\end{eqnarray}
where ${\bf x}_i$ is a $d$-dimensional vector denoting the position of the $i$-th fermion and $r_i^2 = {\bf x}_i \cdot {\bf x}_i$. For convenience, we have set $m = \hbar = 1$. The many-body eigenfunction $\Psi_E({\bf x}_1, \ldots, {\bf x}_N)$ of energy $E$ satisfies the Schr\"odinger
equation 
\begin{eqnarray}\label{Schrod}
\hat H \, \Psi_E({\bf x}_1, \ldots, {\bf x}_N) = E \, \Psi_E({\bf x}_1, \ldots, {\bf x}_N)  \;.
\end{eqnarray}  
Since the many-body Hamiltonian does not contain any interaction term, the many-body wavefunction can be expressed as a product of single particle eigenfunctions. The fermionic constraint however does not allow more than one particle in each single particle state. Consequently, the ground state wavefunction is given by the Slater determinant constructed from the $N$ lowest single particle eigenfunctions 
\begin{eqnarray}\label{psi_0}
\Psi_0({\bf x}_1, \cdots, {\bf x}_N) = \frac{1}{\sqrt{N!}} \det_{1\leq i,j \leq N} \left[ \psi_{{\bf k}_i}({\bf x}_j) \right] \;,
\end{eqnarray}
where ${\bf k}_i$ labels the single particle eigenfunction. In the ground-state, we fill up the single particle levels up to the Fermi energy which will be denoted by $\mu$. For the harmonic potential in $d$ dimensions, the Fermi energy $\mu$ can be computed for large $N$ and is given by~\cite{Castin,dea15b,dea16} 
\begin{eqnarray}\label{expr_mu}
\mu \simeq \left[\Gamma(1+d) \, N\right]^{1/d} \;.
\eea 

%The ground-state energy $E_0$ is just the sum of the single particle energies of all filled levels. 

Knowing exactly the many-body ground-state wavefunction $\Psi_0({\bf x}_1, {\bf x}_2, \cdots, {\bf x}_N)$ given 
in Eq. (\ref{psi_0}), the quantum joint probability density of the positions of the fermions in the ground state is given by  
\begin{eqnarray}\label{jpdf}
P({\bf x}_1, {\bf x}_2, \cdots, {\bf x}_N) = |\Psi_0({\bf x}_1, {\bf x}_2, \cdots, {\bf x}_N)|^2 = \frac{1}{N!} \det_{1\leq i,j \leq N} \left[ \psi^*_{{\bf k}_i}({\bf x}_j) \right] \det_{1\leq i,j \leq N} \left[ \psi_{{\bf k}_i}({\bf x}_j) \right] \;.
\end{eqnarray}
Using $\det(A^T) \det(B) = \det(AB)$, the joint PDF can be re-expressed as a single determinant
\begin{eqnarray}\label{pdf_kernel}
P({\bf x}_1, {\bf x}_2, \cdots, {\bf x}_N)  = \frac{1}{N!} \det_{1\leq i,j \leq N} K_\mu({\bf x}_i,{\bf x}_j) \;,
\end{eqnarray}
where we have defined the kernel $K_\mu({\bf x},{\bf y})$ as
\begin{eqnarray}\label{def_kernel}
K_\mu({\bf x}, {\bf y}) = \sum_{{\bf k}} \theta(\mu-\epsilon_{\bf k}) \psi^*_{\bf k}({\bf x})\psi_{\bf k}({\bf y}) \;.
\end{eqnarray}
This kernel plays a central role for the calculation of the 
correlations. For instance, the $n$-point correlation function $R_n({\bf 
x}_1,\cdots, {\bf x}_n)$ defined as
\begin{eqnarray}\label{eq:def_Rn}
R_n({\bf x}_1,\cdots, {\bf x}_n) = 
\frac{N!}{(N-n)!} \int d{\bf x}_{n+1} \cdots d {\bf x}_N \, 
|\Psi_0({\bf x}_1, \cdots, {\bf x}_N)|^2
\end{eqnarray}
can be expressed as an $n \times n$ determinant 
\begin{eqnarray}\label{eq:Rn_det}
R_n({\bf x}_1,\cdots, {\bf x}_n) = \det_{1\leq i,j \leq n} K_\mu({\bf x}_i,{\bf x}_j) \;.
\end{eqnarray}
This fact demonstrates that the free fermions in the ground state constitute a determinantal process \cite{joh05,bor11} in any dimension. 

A simple consequence of this determinantal structure is that the average local density of fermions can also be expressed very simply in terms 
of the kernel $K_{\mu}({\bf x},{\bf x})$ at identical points. Indeed, the average local density is defined as
\bea\label{R1.2}
%R_1({\bf x}) = N \rho_N({\bf x}) \quad , \quad 
\rho_N({\bf x})  = \frac{1}{N} \Big \langle \sum_{i=1}^N \delta({\bf x}-{\bf x}_i) \Big \rangle_0
\eea 
where $\langle \cdots \rangle_0$ denotes the average w.r.t. the ground state 
quantum probability in Eq. (\ref{psi_0}). It is easy to show that 
\begin{equation}
\rho_N({\bf x})= \frac{1}{N} R_{1}({\bf x}) = \frac{1}{N}\, K_\mu({\bf x},{\bf x})\, .
\label{density_kernel}
\end{equation}
For isotropic quantum potentials, such as the harmonic potential considered here, the density $\rho_N({\bf x}) =  K_{\mu}({\bf x},{\bf x})/N$, or equivalently $K_{\mu}({\bf x},{\bf x})$ only depends on the radial coordinate $r = |{\bf x}|$. Hence
\begin{eqnarray}\label{rho_radial}
\rho_N({\bf x}) = \rho_N(r) = \frac{1}{N} K_\mu(r,r) \;.
\end{eqnarray}
In the limit of large $N$, the average density is isotropic and has a simple form given by \cite{dea15b,dea16}
\bea\label{rho}
\rho_N(r) \simeq \frac{1}{N (2\sqrt{\pi})^d \Gamma(\frac{d}{2}+1)}\, \left(r_{\rm edge}^2 - r^2\right)^{d/2} \theta(r_{\rm edge} - r) \;,
\eea
where $r_{\rm edge}$ is given by \cite{dea15b,dea16}
\bea\label{redge}
r_{\rm edge} = 2^{1/2} \,\left[\Gamma(d+1) \right]^{\frac{1}{2d}} N^{\frac{1}{2d}} = \sqrt{2\,\mu} \;.
\eea
This bulk average density is well known and can also be derived by the standard semi-classical approximation (Thomas-Fermi or Local Density Approximation). However, this bulk form is only valid sufficiently far inside the edge at $r=r_{\rm edge}$. It was pointed out recently that if one zooms in at the edge regime, the density profile does not vanish sharply as in Eq. (\ref{rho}), but instead gets smeared out over a scale $w_N$, due to finite $N$ corrections \cite{dea15b,dea16}. The width of these edge fluctuations was computed explicitly in~\cite{dea15b,dea16}
\begin{eqnarray}\label{wN}
w_N = \frac{1}{\sqrt{2}} \left[ \Gamma(d+1)\,N\right]^{-\frac{1}{6d}} = \frac{1}{\sqrt{2}} \, \mu^{-\frac{1}{6}} \;.
\end{eqnarray}
On this scale, i.e., when $r-r_{\rm edge} \sim w_N$, the smeared density profile is given by $\rho_N(r) = K_\mu(r,r)/N$ where~\cite{dea15b,dea16}
\bea\label{K_Fd}
K_\mu(r,r) = \frac{1}{w_N^d} F_d\left( \frac{r-r_{\rm edge}}{w_N}\right) \;.
\eea
The scaling function $F_d(s)$ was computed explicitly for all $d$ in Ref. \cite{dea15b,dea16}. In this paper, we will need only the large $s$ asymptotics of $F_d(s)$ given by 
\bea\label{Fdz}
F_d(s) \simeq \left( 8 \pi \right)^{-\frac{d+1}{2}} \, s^{-\frac{d+3}{4}} \exp\left(-\frac{4}{3}s^{3/2} \right) \;, \; {\rm as} \;\; s \to \infty \;.
\eea

In this paper, we are interested in the statistics of the position of the farthest fermion from the trap center, in the ground state. More precisely, we define
\begin{eqnarray}\label{rmax}
r_{\max} = \max(r_1, r_2, \ldots, r_N) \;\;\; {\rm where} \;\;\; r_i^2 = {{\bf x_i}}\cdot {\bf x_i} \;, 
\end{eqnarray}
as the radial distance of the farthest fermion. This observable $r_{\rm max}$ is a random variable in
the ground-state, due to the quantum fluctuations encoded in the joint PDF (\ref{jpdf}). Our goal is to compute the probability distribution 
function (PDF) of $r_{\max}$. It is however convenient first to consider the associated cumulative distribution function (CDF)
\begin{eqnarray}\label{cdf}
P(w,N) = {\rm Prob.}(r_{\max} \leq w,N) = {\rm Prob}.[r_1 \leq w, r_2\leq w, \ldots, r_N \leq w] \;.
\end{eqnarray} 
The PDF of $r_{\max}$ is thus given by $p(w,N) = \partial_w  P(w,N)$. To compute $P(w,N)$, it is useful to introduce the indicator function $I_w({\bf x}_i) = 1$ if $r_i \geq w$ and $0$ otherwise, for each fermion. We can then express $P(w,N)$ as
\begin{eqnarray}\label{prod_0}
P(w,N) &=& \Big \langle \prod_{i=1}^N \left(1 - I_w({\bf x}_i)\right)\Big \rangle_0 \;, 
\end{eqnarray}
where $\langle \cdots \rangle_0$ refers to a ground state average with respect to the joint PDF in Eq. (\ref{jpdf}). One can evaluate this average on the right hand side (rhs) of Eq. (\ref{prod_0}) using the following identity valid for any determinantal process 
\bea \label{genf} 
&&  \Big\langle \prod_{i=1}^N (1-f({\bf x}_i)) \Big \rangle_0 = {\rm Det}( I - f\,K_\mu)  \;,
\eea 
where $f({\bf x})$ is an arbitrary function and ${\rm Det}$ denotes a Fredholm determinant, which is interpreted as
\bea\label{Fredholm_id}
{\rm Det}(I - f\,K_\mu) = \exp{\left[\ln {\rm Det}(I-f\,K_\mu)\right]} =  \exp{\left[-\sum_{n=1}^\infty \frac{1}{n} {\rm Tr}\left[(f\, K_\mu)^n\right] \right]}  \;.
\eea
In this trace expansion, the term ${\rm Tr}\left[(f\, K)^n\right]$ is the $n$-dimensional integral
\begin{eqnarray}\label{product}
{\rm Tr}\left[(f\, K)^n\right] = \int d{\bf x}_1 \cdots \int d{\bf x}_n \, f({\bf x}_1)\, K_\mu({\bf x}_1,{\bf x}_2) \, f({\bf x}_2) \, K_\mu({\bf x}_2,{\bf x}_3) \cdots f({\bf x}_n)\, K_\mu({\bf x}_n,{\bf x}_1) \;
\eea
where $K_\mu({\bf x},{\bf y})$ is the kernel defined in (\ref{def_kernel}). Thus, choosing $f({\bf x}) = I_w({\bf x})$ in Eq. (\ref{genf}) gives the identity
\bea\label{hole_proba}
P(w,N) = {\rm Det}(I - I_w K_\mu) = \exp{\left[-\sum_{n=1}^\infty \frac{1}{n} {\rm Tr}\left[(I_w\, K_\mu)^n\right] \right]}
\eea
with 
\bea
{\rm Tr}\left[(I_w\, K_\mu)^n\right] =  \int_{\cal D} d{\bf x}_1 \cdots d{\bf x}_n \, K_\mu({\bf x}_1,{\bf x}_2) \,  K_\mu({\bf x}_2,{\bf x}_3) \cdots K_\mu({\bf x}_n,{\bf x}_1) \; \label{Trn} 
\eea
where the integration domain ${\cal D}$ corresponds to $r_i \geq w$, for each $i=1,2,\cdots, N$. 
The reader should note that, up to this point, the formulas are valid for arbitrary finite $N$.
In the next section we will use Eq. (\ref{hole_proba}) to 
analyze the limiting form of $P(w,N)$ for large $N$. 

\subsection{Limiting distribution of $r_{\max}$ for large $N$ in $d \geq 2$}
\label{sec:limit} 

Our goal is to compute the limiting form of $P(w,N)$ for large $N$ in all dimensions. As discussed in the introduction (see also later in section \ref{sec:angular}), the case $d=1$ is special, where the limiting distribution in the large $N$ limit, suitably centered and scaled, is given by a product of two Tracy-Widom distributions (see section \ref{sec:angular} for details). In this subsection, we focus on $d \geq 2$ and we will see that the results here are very different from the $d=1$ case. 

Our starting point is the expression for $P(w,N)$ in Eq. (\ref{hole_proba}). For large $N$, we anticipate that the PDF $p(w,N) = \partial_w P(w,N)$ must be peaked around the average value of $r_{\max}$. This average value, to leading order for large $N$, coincides with the edge of the density 
$r_{\rm edge}$ in Eqs. (\ref{rho}) and (\ref{redge}). Since $r_{\rm edge}$ is large for large $N$, we need to analyze the expression for $P(w,N)$ in Eq. (\ref{hole_proba}) when $N$ and $w$ are both large. For large $w$ and $d \geq 2$, it turns out that $P(w,N)$ in Eq. (\ref{hole_proba}) is dominated by the $n=1$ term of the trace expansion -- the proof of this fact is given in Appendix \ref{sec:higher}. Keeping only this $n=1$ term in (\ref{hole_proba}) gives
\begin{equation}
P(w,N)\simeq \exp\left[- {\rm Tr}\left(I_w\, K_\mu\, \right)\right]
= \exp\left[-\int_{\cal D } K_\mu({\bf x},{\bf x})\, d{\bf x}\right]
= \exp\left[-S_d\, \int_w^{\infty} K_\mu(r,r)\, r^{d-1}\, dr\right] \;,
\label{fred.2}
\end{equation}
where $S_d= 2\pi^{d/2}/\Gamma[d/2]$ is the surface area of a unit sphere in $d$ dimensions. Since $w$ is close to $r_{\rm edge}$, we can use the scaling form in Eq. (\ref{K_Fd}) for $K_{\mu}(r,r)$. 
Using Eq.~(\ref{K_Fd}) then gives
%
%
%the scaling form of $K_\mu(r,r)$ from Eq. (\ref{kernel.1}) in
%the integral on the rhs of Eq. (\ref{fred.2}), 
%and using $\frac{r_{\rm edge}}{w_N}= 2\,\mu^{2/3}$ (from Eq. 
%(\ref{para.1})), one gets to leading order
\bea\label{P_1}
P(w,N) \simeq \exp{\left[ -S_d\, \left(\frac{r_{\rm edge}}{w_N}\right)^{d-1} \int_{\tilde w}^{\infty} F_d(s) \, ds \right]} \;,
\eea
where
\begin{equation}
\tilde w= \frac{w-r_{\rm edge}}{w_N}= (w-\sqrt{2\mu})\, \sqrt{2}\, 
\mu^{1/6}\, .
\label{scaled_w}
\end{equation} 
Note that in Eq. (\ref{scaled_w}) we have used the expressions for $r_{\rm edge}$ and $w_N$, in terms of $\mu$, given respectively in Eq.~(\ref{redge}) and (\ref{wN}). Using further that $r_{\rm edge}/w_N = \, 2 \, \mu^{2/3}$ in Eq. (\ref{P_1}) gives
\begin{equation}
P(w,N) \simeq \exp\left[- S_d\, 2^{d-1}\, \mu^{2(d-1)/3}\, \int_{\tilde 
w}^{\infty} F_d(s)\, ds \right] \,.
\label{fred.3}
\end{equation}
For large $\mu$ and $d\geq 2$, the term $\mu^{2(d-1)/3}$ is large and consequently $P(w,N)$ is non zero only when the integral $ \int_{\tilde 
w}^{\infty} F_d(s)\, ds $ is small and of order $\mu^{-2(d-1)/3}$. For this integral to be small, $\tilde w$ must be large. Hence we can evaluate this integral using the asymptotic form of $F_d(s)$ given in Eq. (\ref{Fdz}). Keeping only the leading term gives
\begin{equation}
I(\tilde w)= \int_{\tilde w}^{\infty} F_d(s)\, ds\simeq \frac{1}{2\, 
(8\pi)^{(d+1)/2}}\, \frac{1}{ {\tilde w}^{(d+5)/4}}\, e^{-\frac{4}{3}\, 
{\tilde w}^{3/2}} \;.
\label{Fdz.2}
\end{equation}
Substituting this result in Eq. (\ref{fred.3}) and exponentiating the prefactors inside the exponential, we 
get
\begin{equation}
P(w, N) \simeq \exp\left[- B_d\,\, e^{-\frac{4}{3} {\tilde w}^{3/2} + 
\frac{2(d-1)}{3} \ln \mu - \frac{(d+5)}{4}\, \ln (\tilde w)} \right] \,,
\label{fred.4}
\end{equation}
where $B_d$ is given by
\bea\label{Bd}
B_d= S_d\, \frac{2^{d-2}}{(8 \pi)^{(d+1)/2}} = \frac{2^{-\frac{d+5}{2}}}{\Gamma(d/2) \sqrt{\pi}} \;.
\eea

To obtain the limiting form of $P(w,N)$ in Eq. (\ref{fred.4}), we need to further center and suitably scale the variable $\tilde w$. For this purpose, we set
\begin{equation}
{\tilde w}= a_\mu + b_{\mu}\, z
\label{ab.1}
\end{equation}
where the two unknowns $a_\mu$ (the centering parameter or location) and $b_\mu$ (the scale parameter) will be chosen such that $P(w,N)$ becomes independent of $\mu$ (for large $\mu$) and is only a function of the variable $z$. We further anticipate (to be checked a posteriori ) that, for large $\mu$, $a_\mu$ is large and $b_\mu$ is small, while $z$ is fixed ($\sim {\cal O}(1)$). Substituting this relation (\ref{ab.1}) in Eq.~(\ref{fred.4}), expanding for large $\mu$, and keeping 
leading order terms, one gets
\begin{equation}
P(w, N) \simeq \exp\left[- B_d\,\, e^{- \frac{4}{3}\, a_\mu^{3/2}+ 
\frac{2(d-1)}{3}\, \ln \mu - \frac{(d+5)}{4}\, \ln (a_\mu) - 2\, 
\sqrt{a_{\mu}}\, b_{\mu}\, z}\right] \;.
\label{fred.5}
\end{equation}
Hence we choose 
\begin{equation}
b_\mu =  \frac{1}{2 \sqrt{a_\mu}}\,  
\label{ab.2}
\end{equation}
and $a_\mu$ to be such that
\begin{equation}
B_d\,\, e^{- \frac{4}{3}\, a_\mu^{3/2}+
\frac{2(d-1)}{3}\, \ln \mu - \frac{(d+5)}{4}\, \ln (a_\mu)}=1 \;.
\label{ab.3}
\end{equation}
This then gives the limiting $N$-independent distribution 
\bea
P(w,N) \simeq \exp{[-e^{-z}]} \;. \label{Gumbel0}
\eea
Using (\ref{scaled_w}), the result in Eq. (\ref{Gumbel0}) translates to
\begin{equation}
\tilde w = (w-\sqrt{2\mu})\, \sqrt{2}\, \mu^{1/6}= a_\mu + 
\frac{1}{2\,\sqrt{a_\mu}}\, z \;,
\label{fred.6}
\end{equation}
where the random variable $z$ has the Gumbel CDF as in (\ref{Gumbel0}). Hence, in the large
$N$ limit, one 
can express the original random variable $r_{\max}$ as an identity in law
\begin{equation}
r_{\max}= \sqrt{2\mu} + \frac{1}{\mu^{1/6}}\, \left[\frac{1}{\sqrt{2}}\, a_\mu
+ \frac{1}{2\sqrt{2\, a_\mu}}\, Z_G\right]\,,
\label{fred.7}
\end{equation}
where ${\rm Prob.}[Z_G\le z]= \exp[-e^{-z}]$. The parameter $a_\mu$ is determined from Eq. (\ref{ab.3}) in the limit of large $\mu$. Solving  Eq. (\ref{ab.3}) to leading orders for large $\mu$, we get 
\bea\label{amu}
a_\mu \simeq \left(\frac{d-1}{2} \ln \mu \right)^{\frac{2}{3}}  + \frac{1}{2^{2/3} \left[ (d-1) \ln \mu\right]^{1/3}} \, \ln{\left[\frac{B_d}{(\frac{d-1}{2} \ln \mu)^{\frac{d+5}{6}}} \right]} \;,
\eea
where $B_d$ is given in Eq. (\ref{Bd}). Equation (\ref{fred.7}), together with (\ref{amu}), is the main new result of this paper. Clearly, this Gumbel form for the limiting distribution holds strictly for $d > 1$.

\section{Interpretation of the Gumbel law}\label{sec:angular}

In the previous section we have investigated the statistics of $r_{\max} = \max \{r_1, r_2, \cdots, r_N\}$ of $N$ non-interacting fermions in the ground state of a $d$-dimensional harmonic oscillator with $d>1$. As discussed before, even though the fermions are non-interacting, the Pauli exclusion principle induces strong effective (repulsive) correlations between them. This makes the positions $r_i$'s of the fermions in the ground state strongly correlated. Therefore studying the statistics of $r_{\max}$ corresponds to studying the extreme value statistics of a strongly correlated set of random variables. For large $N$, we found that the limiting distribution of $r_{\max}$, suitably centered and scaled as in Eq. (\ref{fred.7}), is given by the Gumbel distribution. It is well known that the Gumbel law also appears in the context of classical extreme value statistics of {\it independent} and identically distributed (i.i.d.) random variables. So the fact that the Gumbel law also appears in a strongly correlated system of fermions in the ground state of a $d$-dimensional oscillator, with $d>1$, is rather surprising. Is there any mechanism by which the effective random variables become ``decorrelated'' such that the Gumbel distribution emerges naturally? In the derivation presented in the previous section, this mechanism is not manifest. 

In the present section, we present an alternative derivation of the same results as in Eqs. (\ref{fred.7}) and (\ref{amu}) using a completely different method that demonstrates explicitly how such a ``decorrelating'' mechanism emerges. The main idea is to decompose the 
ground state wave function in radial and angular sectors. For each angular quantum number, one effectively obtains a one-dimensional problem for a certain number of non-interacting fermions, in an effective quantum potential characterized by the angular quantum number, much like the hydrogen atom. In each of these angular sectors, the distribution of the farthest fermion position is non trivial. However, different angular sectors decouple and effectively one has to look at the maximum of a collection of independent, but non-identically distributed random variables. We will see below that indeed this mechanism eventually leads to the same Gumbel law, as in Eqs. (\ref{fred.7}) and (\ref{amu}).       

To present this mechanism in a clear fashion, we first remind the reader how the single particle wavefunction can be decomposed in radial and
angular sectors for a $d$-dimensional harmonic oscillator. Next we consider the $N$-particle problem and show that the CDF $P(w,N)$ of $r_{\max}$ factorizes into a product of determinants, each corresponding to a different angular quantum number. Each of these determinants can be mapped 
onto the distribution of the largest eigenvalue of a Laguerre-Wishart matrix. Finally, using the known results for the latter problem, we arrive at the 
result in Eqs. (\ref{fred.7}) and (\ref{amu}).

\subsection{Single particle in a $d$-dimensional harmonic potential}

We first recall the well known results for the quantum mechanics of a single particle in a $d-$dimensional
harmonic potential. The Hamiltonian of the particle is given by:
\begin{eqnarray}\label{def_h_single}
\hat h = -\frac{1}{2} \nabla_{\bf x}^2 + \frac{1}{2} r^2 \;,  
\end{eqnarray}
where ${\bf x}$ is a $d$-dimensional vector denoting the position of the particle and $r^2 = {\bf x} \cdot {\bf x}$. The single particle eigenstates $\psi_{\bf k}({\bf x})$ satisfy the Schr\"odinger equation
\begin{eqnarray}
\hat h \, \psi_{\bf k}(\bf x) = \epsilon_{\bf k}\, \psi_{\bf k}(\bf x) \;, \label{eigen_single}
\end{eqnarray}
where $\epsilon_{\bf k}$'s denote the single particle discrete energy levels. Note that we just use ${\bf k}$ to label the single particle eigenstates (which are not plane waves). While it is easy to write down explicitly the eigenfunctions
and eigenvalues in Cartesian coordinates, it is more convenient, for our purpose, to present the solution in spherical coordinates. The Hamiltonian operator can be split into a radial and an angular part. We denote the wave function in spherical coordinates by $\psi_{\bf k}(r,{\boldsymbol{\theta}})$ where ${\boldsymbol \theta}$ is 
a $(d-1)$-dimensional angular vector. In these coordinates the Schr\"odinger equation~(\ref{eigen_single}) becomes
\begin{eqnarray}
\frac{1}{2} \left[-\frac{1}{r^{d-1}} \frac{\partial}{\partial r} \left(r^{d-1} \frac{\partial}{\partial r} \right) - \frac{\hat\Delta_{\boldsymbol \theta}}{r^2} + r^2 \right] \psi_{\bf k}(r,{\boldsymbol \theta}) = \epsilon_{\bf k} \, \psi_{\bf k}(r,{\boldsymbol \theta}) \;,
\end{eqnarray} 
where $\hat \Delta_{\boldsymbol \theta}$ is the Laplacian on the surface of  a $d$-dimensional unit sphere \cite{mosh96}. Since the potential 
depends only on the radial coordinate $r = |{\bf x}|$, we can write the solution using separation of variables as $\psi_{\bf k}(r,{\boldsymbol \theta}) = \phi(r)Y_{\bf L}({\boldsymbol \theta})$, where ${\bf L}$ stands collectively for all the angular quantum numbers. For simplicity, we suppress, for the moment, the explicit quantum number dependence (i.e., ${\bf k}$ dependence) of the eigenfunction $\phi(r)$.
%\begin{equation}\label{separate}
%\psi_{n, {\bf l}}(r,{\boldsymbol \theta}) = \phi(r)Y_{\bf l}({\boldsymbol \theta}) \;,
%\end{equation}
%\frac{1}{r^{\frac{d-1}{2}}}\chi_{n,{\bf l}}(r)
The angular part $Y_{\bf L}({\boldsymbol \theta})$ satisfies the eigenvalue equation
\begin{eqnarray}\label{eigen_Y}
\hat \Delta_{\boldsymbol \theta} Y_{\bf L}({\boldsymbol \theta}) = \lambda \, Y_{\bf L}({\boldsymbol \theta}) \;.
\end{eqnarray}
It turns out that, while the eigenfunctions $Y_{\bf L}({\boldsymbol \theta})$ depend explicitly on all the angular quantum numbers denoted as ${\bf L}$, the eigenvalues, in contrast, are labelled just by one single scalar quantum number \cite{mosh96}, $\lambda = - l(l+d-2)$ where $l = 0, 1, 2 \ldots$. Each $l$-sector ($l > 0$) is degenerate and the degeneracy is given by \cite{mosh96}
\begin{equation}\label{degeneracy}
g_d(l) = {(2l+d-2)(l+d-3)!\over l! (d-2)!} = \frac{(2l+d-2) \Gamma(l+d-2)}{\Gamma(l+1) \Gamma(d-1)} \;.
\end{equation}
Note that $g_d(0) = 1$ in all dimensions $d \geq 1$. In $d=1$, it follows from Eq. (\ref{degeneracy}) that 
\begin{eqnarray}\label{gd1}
g_1(0) =   1 \;\;, \;\; g_1(1) = 1 \;\;, \;\; g_1(l) = 0 \;\; {\rm for \; all} \;\; l \geq 2 \;.
\end{eqnarray}
This is consistent with the fact that in $d=1$, the quantum number $l$ has only two allowed values $l = 0$ and $l=1$, corresponding to the angles $\theta = 0$ and $\theta = \pi$ respectively.  

These angular eigenfunctions in Eq. (\ref{eigen_Y}) satisfy the orthogonality condition
\begin{equation}\label{ortho}
\int d{\boldsymbol \theta}\ 
  Y_{\bf L}({\boldsymbol \theta})Y_{{\bf L}'}({\boldsymbol \theta}) = \delta_{{\bf L},{\bf L}'} \;.
\end{equation}
For a fixed $l$ the radial part satisfies the eigenvalue equation
\begin{eqnarray}\label{eq:phi}
\left[-\frac{1}{2} \frac{d^2}{d r^2 } - \frac{d-1}{2\,r} \frac{d}{dr} + \frac{l(l+d-2)}{2\,r^2} + \frac{r^2}{2} \right] \phi(r) = \epsilon_{\bf k} \phi(r) \;.
\end{eqnarray}
This equation for $\phi(r)$ (\ref{eq:phi}) can be reduced to a standard one-dimensional Schr\"odinger equation by setting $\phi(r) = \chi(r)/r^{(d-1)/2}$ that eliminates the first order $d/dr$ term in Eq. (\ref{eq:phi}) and we get
\begin{equation}\label{eq:chi}
 -{1\over 2} {d^2 \chi(r)\over dr^2} +{1\over 2}\left[ r^2 + {( l+ {d-3\over 2})(l + {d-1\over 2})\over r^2}\right] \chi(r) = \epsilon_{\bf k} \, \chi(r) \;\;\; , \;\;\;r \geq 0 \;.
\end{equation}
This $1d$ Schr\"odinger equation is solvable with discrete energy levels labelled by $n$ and $l$ \cite{mosh96} (see Fig. \ref{Fig_levels})
\begin{eqnarray}\label{epsilon_k}
\epsilon_{\bf k} \equiv \epsilon_{n, l} = 2n + l + \frac{d}{2} \;,
\end{eqnarray}
 where $n=0, 1, 2, \ldots$. The associated radial eigenfunctions $\chi_{n,l}(r)$ are given explicitly by
 \begin{eqnarray}\label{expr_chi}
 \chi_{n, l}(r) = A_{n,l} \; r^{l+{d-1\over 2}} \; L_{n}^{l+{d-2\over 2}}\left(r^2\right) \;e^{-\frac{r^2}{2}} \;,
 \end{eqnarray}
where $L_{n}^{\alpha}$ are generalized Laguerre polynomials (which are of degree $n$) and $A_{n,l}$ are normalization constants, such that the $\chi_{n,l}$ satisfy the orthogonality relation 
 \begin{equation}
\int_0^\infty  dr\ \chi_{n,l}(r)\chi_{n',l}(r)=\delta_{n,n'}.
\end{equation}

Thus, finally, the eigenfunctions $\psi_{\bf k}({\bf x})$ of the $d$-dimensional harmonic oscillator, labelled by the quantum numbers ($n,{\bf L}$), are given by 
\begin{eqnarray}\label{psi_final}
\psi_{n, l}(r,{\boldsymbol \theta}) = \frac{1}{r^{\frac{d-1}{2}}} \, \chi_{n,l}(r) \, Y_{\bf L}({\boldsymbol \theta}) \;,
\end{eqnarray}
with an associated eigenvalue $\epsilon_{n,l} = (2n + l + d/2)$. 

Note that in $d=1$, since only $l=0$ and $l=1$ are allowed, the additional potential term $l(l-1)/r^2$ in Eq. (\ref{eq:chi}) vanishes
and $\chi_{n,l}(r)$ satisfies the $1d$ Schr\"odinger equation in a harmonic potential, but restricted only to positive half-space ($r \geq 0$). 
The energy levels in Eq. (\ref{epsilon_k}) get divided into two sectors labeled by $l=0$ and $l=1$
\begin{eqnarray}\label{epsilon_k_1d}
\epsilon_{n,0} = 2n + \frac{1}{2} \;\;\;, \;\;\; \epsilon_{n,1} = (2n+1) + \frac{1}{2} \;.
\eea 
One sees that the $l=0$ sector corresponds to the even eigenfunctions of the $1d$ harmonic oscillator in the full space, while the $l=1$
sector corresponds to the odd eigenfunctions.

\begin{figure}
\includegraphics[width=0.7\linewidth]{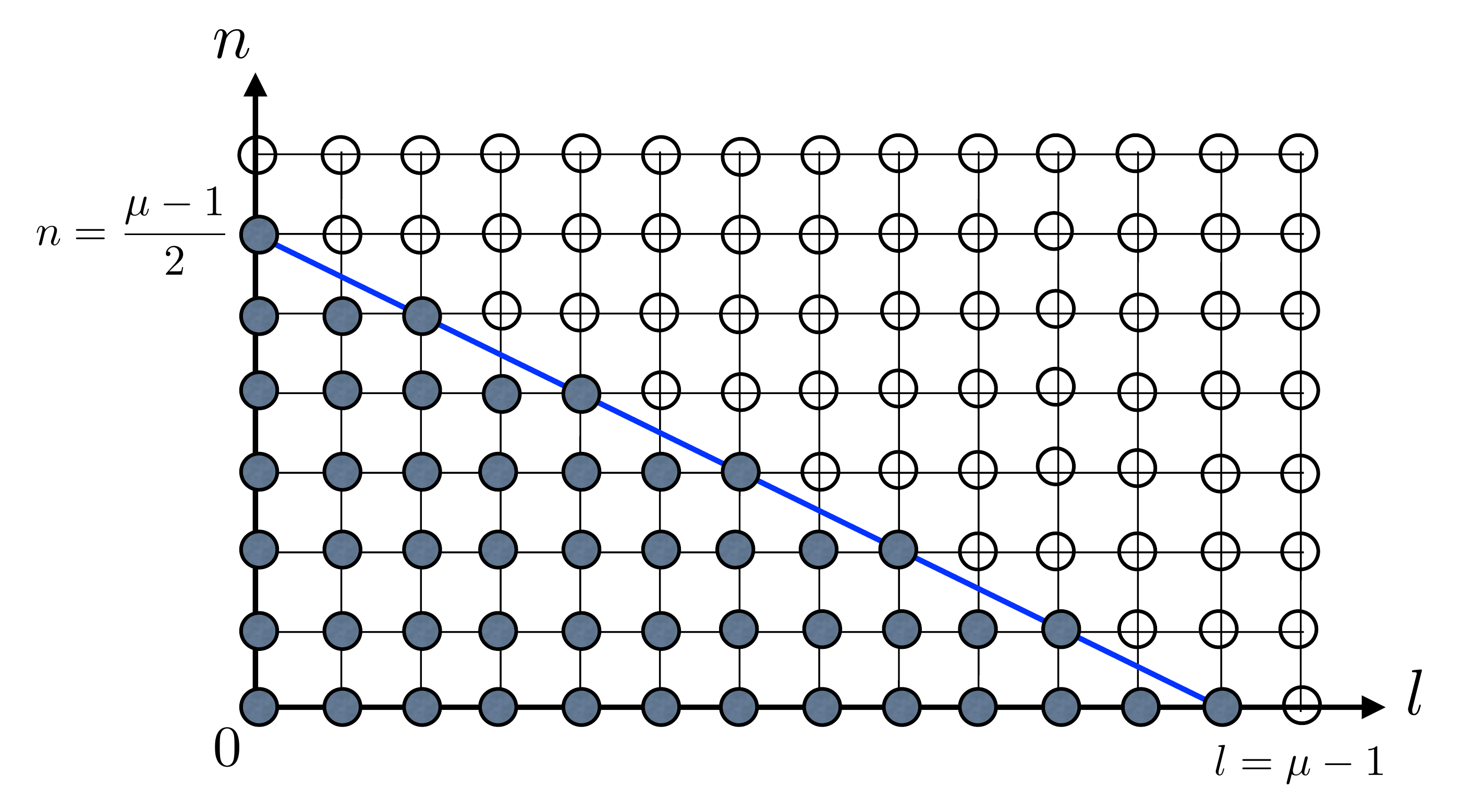}
\caption{Occupation of the energy levels in the plane $(l,n)$ [see Eq. (\ref{psi_final})] for  $N=98$ spin-less fermions in the ground state of the two-dimensional ($d=2$) isotropic harmonic oscillator. The filled circles indicate occupied states while the open circles are unoccupied. In this case the energy levels are $\epsilon_{n,l} = 2n+l+1$ and the degeneracy is $g_{n,l} = 2$ [see Eq. (\ref{degeneracy})], corresponding to 
${Y}_{\bf L}(\theta) = e^{ \pm i l \theta}/\sqrt{2\pi}$, with $l=0,1, \cdots$. Thus each dot actually corresponds to $2$ distinct quantum states which implies that the Fermi energy (corresponding to the levels connected by the solid blue line) is $\mu = 13$. Note that the last energy level at the Fermi energy $\mu$ is fully occupied, the situation to which we restrict ourselves here (see text).}\label{Fig_levels}
\end{figure}

\subsection{Many-body ground-state for $N$ free fermions in a $d$-dimensional harmonic potential}

\noindent{\it Ground-state wave function.} We now consider $N$ spin-less free fermions in a $d$-dimensional harmonic potential. The many-body Hamiltonian
is given by
\begin{eqnarray}\label{def_H}
\hat H = \sum_{i=1}^N \hat h_i \;, \; \;\;\; {\rm where} \;\;\;\; \hat h_i = -\frac{1}{2}\nabla_{{\bf x}_i}^2 + \frac{1}{2} r_i^2 \;, 
\end{eqnarray}
where ${\bf x}_i$ is a $d$-dimensional vector denoting the position of the $i$-th fermion and $r_i^2 = {\bf x}_i \cdot {\bf x}_i$. The many-body eigenfunction $\Psi_E({\bf x}_1, \ldots, {\bf x}_N)$ of energy $E$ satisfies the Schr\"odinger
equation 
\begin{eqnarray}\label{Schrod}
\hat H \, \Psi_E({\bf x}_1, \ldots, {\bf x}_N) = E \, \Psi_E({\bf x}_1, \ldots, {\bf x}_N)  \;.
\end{eqnarray}  
Each single particle Hamiltonian $\hat h_i$ has a discrete set of eigenvalues $\epsilon_{n_i, l_i} = (2 n_i + l_i + d/2)$ as given above (\ref{epsilon_k}). Since the many-body Hamiltonian does not contain any interaction term, the many-body wavefunction can be expressed as a product of single particle eigenfunctions. The fermionic constraint however does not allow more than one particle in each single particle state. Consequently, the ground state wavefunction is given by the Slater determinant constructed from the $N$ lowest single particle eigenfunctions 
\begin{eqnarray}\label{psi_0_2}
\Psi_0({\bf x_1}, \cdots, {\bf x_N}) = \frac{1}{\sqrt{N!}} \det_{1\leq i,j \leq N} \left[ \psi_{{\bf k}_i}({\bf x}_j) \right] \;,
\end{eqnarray}
where ${\bf k}_i = (n_i, {\bf L}_i)$ labels the single particle eigenfunction. In the ground-state, we fill up the single particle levels up to the Fermi energy which will be denoted by $\mu$. The ground-state energy $E_0$ is just the sum of the single particle energies of all filled levels. 

Note that in $d>1$ there is a large degeneracy of the single particle energy levels,
i.e., many different couples $(n,l)$ have the same value of $\epsilon_{n, l}$, and in
addition, for each such couple there is an angular degeneracy $g_d(l)$. 
In this work we will restrict ourselves to the case where the set of degenerate 
single particle energy levels with highest energy is fully occupied. In this
case the $N$- body ground state is unique, and it is the case studied here (we refer the reader to Ref. \cite{dea16} for further discussions of the degenerate ground states).

\vspace*{0.5cm}

\noindent{\it Fermi energy}. Knowing explicitly the single particle energy levels $\epsilon_{n,l} = (2n+l+d/2 )$, together with their degeneracies $g_d(l)$ given in Eq. (\ref{degeneracy}), one can readily 
relate the number of fermions $N$ to the 
%compute the 
Fermi energy $\mu$ as follows (see Fig. \ref{Fig_levels})
\begin{eqnarray}\label{mu_1}
N = \sum_{l,n \geq 0} g_d(l) \theta(\mu - \epsilon_{n,l}) \;.
\end{eqnarray} 
%Note that in Eq. (\ref{mu_1}) we have assumed that the Fermi level (i.e., the last level) is fully occupied. 
Carrying out the sum over $n$, one obtains
\begin{eqnarray}\label{mu_2}
N = \sum_{l\geq 0} g_d(l) \, \left(1 + {\rm int}\left(\frac{\mu-l-d/2}{2} \right) \right) \theta\left(\mu-l-d/2 \geq 0 \right)\;,
\end{eqnarray}
where ${\rm int}{(x)}$ denotes the integer part of $x$. We can easily estimate $\mu$ for large $N$. For this, one needs to distinguish the $d=1$ case and $d \geq 2$. 

\vspace*{0.5cm}

\noindent 
{\bf Case $d=1$}. In $d=1$, the sum over $l$ in Eq. (\ref{mu_2}) runs over two terms, $l=0$ and $l=1$, with $g_1(0) = g_1(1) = 1$. It follows immediately from Eq. (\ref{mu_2}) that, for large $N$,
\bea\label{mu_1d}
\mu \simeq N \;.
\eea

\vspace*{0.5cm}
\noindent 
{\bf Case $d\geq 2$}. In the large $N$ limit, $\mu$ is expected to be large. In this case, the sum on the right hand side of Eq. (\ref{mu_2}) is dominated by large $l$. For large $l$ and $d \geq 2$, using Stirling's formula in Eq. (\ref{degeneracy}), the degeneracy $g_d(l)$ behaves asymptotically as, 
\bea\label{gd}
g_d(l) \simeq \frac{2\, l^{d-2}}{\Gamma(d-1)} \;. 
\eea
Substituting this behavior in the sum and rescaling $l = \mu \ell$, one can replace, for large $\mu$, the discrete sum by an integral over $\ell \in [0,1]$.  In this large $\mu$ limit, one can also replace ${\rm int}((\mu - l -d/2)/2)$ by $\mu(1 - \ell)/2$. This gives
\begin{equation}\label{mu_3}
N \simeq\frac{\mu^d}{\Gamma(d-1)}\int_0^1 (1-\ell)\ell^{d-2} \, dz = \frac{\mu^d}{\Gamma(d+1)} \;.
\end{equation} 
One obtains
\begin{eqnarray}\label{mu_4}
\mu \simeq \left( \Gamma(d+1) \, N\right)^{1/d} \;,
\end{eqnarray}
thus recovering the result mentioned in (\ref{expr_mu}). Interestingly, although the derivation holds only for $d > 1$, the result in Eq. (\ref{mu_4}) remains valid even for $d=1$, where it gives the correct result $\mu \simeq N$ in (\ref{mu_1d}). 

Finally, we introduce another quantity that will be useful later. For a given $\mu$, it is clear from Eq. (\ref{mu_1}), that for fixed $l$ the sum over $n$ is cut off at the top due to the theta function (see also Fig. \ref{Fig_levels}). 
Let $N_l(\mu)$ be this maximal allowed value. It is given by 
\begin{eqnarray}\label{Nl}
N_l(\mu) = {\rm int} \left(\frac{\mu-l-d/2}{2} \right) \;.
\end{eqnarray}
For $d=1$, there are only two allowed values $l=0$ and $l=1$ and hence for large $\mu$, it follows from Eq. (\ref{Nl}) that these two sectors have
the same number of states
\bea\label{Nl_1d}
N_{l=0}(\mu) \simeq \frac{\mu}{2} \simeq \frac{N}{2} \;\;\;,   \;\;\;N_{l=1}(\mu) \simeq \frac{\mu}{2}  \simeq \frac{N}{2} \;.
\eea

\subsection{The distribution of the position of the farthest fermion}

We now turn to the CDF $P(w,N)$ of $r_{\max}$ defined in Eq. (\ref{cdf}). Using the representation in Eq. (\ref{prod_0}) and the joint PDF in Eq. (\ref{jpdf}), we get
\begin{eqnarray}\label{prod}
P(w,N) &=& \Big \langle \prod_{i=1}^N \left(1 - I_w({\bf x}_i)\right)\Big \rangle_0 \\
&=& \frac{1}{N!} \int d{\bf x}_1 \ldots \int d{\bf x}_N \, \det_{1\leq i,j \leq N} \left[ \psi^*_{{\bf k}_i}({\bf x}_j) \right] \det_{1\leq i,j \leq N} \left[ \psi_{{\bf k}_i}({\bf x}_j) \right] \,  \prod_{i=1}^N \left(1 - I_w({\bf x}_i)\right) \;.
\end{eqnarray}
In section II, we had seen that $P(w,N)$ can be expressed as a Fredholm determinant [see Eq. (\ref{hole_proba})]. Here, we show that it can be expressed as a determinant of a finite matrix (as opposed to the functional  or Fredholm
determinant), using the powerful Andreief-Cauchy-Binet identity. According to this identity, one has 
\begin{eqnarray}\label{cauchy_binet}
\int d{\bf x}_1 \ldots \int d{\bf x}_N \, \prod_{i=1}^N h({\bf x}_i) \det_{1\leq i,j \leq N} \left[f_i({\bf x}_j) \right] \det_{1\leq i,j \leq N} \left[g_i({\bf x}_j) \right] = N ! \det_{1 \leq i,j \leq N} \left[ \int d{\bf x}\, h({\bf x}) f_i({\bf x}) g_j({\bf x}) \right] \;,
\end{eqnarray}
where $f_i({\bf x})$, $g_j({\bf x})$ and $h({\bf x})$ are arbitrary integrable functions. Choosing $h({\bf x}) = 1 - I_w({\bf x})$ and setting $f_i = \psi_{{\bf k}_i}$ and $g_j = \psi^*_{{\bf k}_j}$, we immediately get
\begin{eqnarray}\label{cdf_det}
P(w,N) &=& \det_{1 \leq i,j \leq N} \left[ \int d{\bf x}\, \left[1 - I_w({\bf x})\right] \, \psi^*_{{\bf k}_i}({\bf x})  \, \psi_{{\bf k}_j}({\bf x}) \right] =  \det_{1 \leq i,j \leq N} \left[\delta_{ij} - \int d{\bf x}\, I_w({\bf x}) \, \psi^*_{{\bf k}_i}({\bf x})  \, \psi_{{\bf k}_j}({\bf x}) \right] \;,
\end{eqnarray}
where we have used the orthogonality of $\psi_{\bf k}$'s in the second equality. Using the parametrization of the quantum numbers ${\bf k} = (n,{\bf L})$, the expression above can be rewritten as
\begin{eqnarray}\label{cdf_det_2}
P(w,N) &=&  \det_{1 \leq i,j \leq N} \left[\delta_{ij} - \int_{r \geq w} d{\bf x} \, \psi^*_{n_i, {\bf L}_i}({\bf x}) \psi_{n_j,{\bf L}_j}({\bf x})   \right] \;\;, \; \; {\rm where} \;\; r^2 = {\bf x}\cdot {\bf x} \;,
\end{eqnarray} 
where $\psi_{n,{\bf L}}({\bf x})$ is given in Eq. (\ref{psi_final}). Using $d{\bf x} = r^{d-1} \,dr \, d{\boldsymbol \theta}$ in the integral on the rhs of Eq. (\ref{cdf_det_2}) and the explicit form of $\psi_{n,{\bf L}}({\bf x})$ in Eq. (\ref{psi_final}), we see that the factor $r^{d-1}$ cancels.  
Carrying out the angular integral and using the orthogonality condition in Eq.~(\ref{ortho}),
one obtains
\begin{eqnarray}\label{integral_1}
\int_{r \geq w} d{\bf x} \, \psi^*_{n_i, {\bf L}_i}({\bf x}) \psi_{n_j,{\bf L}_j}({\bf x}) = \delta_{{\bf L}_i, {\bf L}_j} \int_{w}^{\infty} dr \, \chi_{n_i,l_i}(r) \chi_{n_j,l_i}(r)  \;.
\end{eqnarray}  
Using this result in Eq. (\ref{integral_1}), we see that the determinant on the right hand side  of Eq. (\ref{cdf_det_2}) decouples into a product of determinants, each corresponding to a particular angular sector. Taking into account the degeneracy $g_d(l)$ in Eq.~(\ref{degeneracy}) associated to each value of the quantum number $l$, we get
\begin{eqnarray}\label{product_det}
P(w,N) = \prod_{l \geq 0}^{ {\rm int}(\mu-\frac{d}{2})} \, \Big[P_l(w,m_l) \Big]^{g_d(l)} \;,
\end{eqnarray} 
where, for $d>1$,  
\bea\label{def_m}
m_l = N_l(\mu) + 1 \;\; {\rm with} \;\; N_l(\mu) =  {\rm int} \left(\frac{\mu-l-d/2}{2} \right) \;,
\eea
denotes the number of fermions in the states of angular momentum $l$. The product over $l$ in Eq. (\ref{product_det}) runs up to $\mu-d/2$. 
By contrast, for $d=1$, the product in Eq. (\ref{product_det}) runs only over two values $l=0$ and $l=1$, with $g_1(0) = g_1(1) = 1$ and the corresponding values of $m_l$ are $N/2$ for each sector [see Eq. (\ref{Nl_1d})].

In Eq. (\ref{product_det}), $P_l(w,m_l)$ is an $m_l \times m_l$ determinant given by
\begin{eqnarray}\label{P_l}
P_l(w,m_l) = \det_{1 \leq i,j \leq m_l} \left[ \delta_{i,j} - \int_{w}^\infty \chi_{n_i,l}(r) \chi_{n_j,l}(r) \, dr\right] \;,
\end{eqnarray}
where $n_i = 0, 1, \ldots, m_l-1$. In Eq. (\ref{P_l}), the radial wavefunction $\chi_{n,l}(r)$ is given explicitly in Eq.~(\ref{expr_chi}). We will see below that this determinant in Eq.~(\ref{P_l}) can actually be interpreted as the cumulative distribution of the position of the rightmost fermion among a set of $m_l$ spin-less fermions in a one-dimensional quantum problem. 

\vspace*{0.5cm}

\noindent{\it The effective $1d$ problem.} In order to interpret $P_l(w,m_l)$ in (\ref{P_l}), we consider an effective quantum 
$1d$ model on the positive real axis with a single particle Hamiltonian given by
\begin{eqnarray}
\hat h_{\rm eff} = -{1\over 2} {d^2\over dr^2} +{1\over 2}\left[ r^2 + \frac{\alpha(\alpha-1)}{r^2}\right],\label{heff}
\end{eqnarray}
with $\alpha = l + (d-1)/2$. The single particle eigenfunctions are given by $\chi_{n,l}(r)$ given in Eq.~(\ref{expr_chi}) with eigenvalues 
$\epsilon_{n,l} = (2n+l+d/2)$ as in Eq. (\ref{epsilon_k}). We now consider $m_l = N_l(\mu)+1$ non-interacting fermions with many-body
Hamiltonian which is just the sum of the single particle Hamiltonians in Eq. (\ref{heff}). Indeed, this non-interacting fermion problem has
already been studied in the context of non-intersecting fluctuating interfaces in $1+1$ dimensions \cite{nad09} and a relation to Laguerre-Wishart 
random matrices was found, as we recall below.  
 
The many-body ground state wavefunction 
of this effective $1d$ fermion model can be written again as a Slater determinant of the first $m_l$ single particle eigenstates
\begin{eqnarray}
\Psi_{{\alpha}}(r_1,r_2,\cdots\,, r_{m_l}) = {1\over \sqrt{Z_{m_l,\alpha}}}\left[\prod_{i=1}^{m_l} r_i^{\alpha} \, e^{-\frac{1}{2}r_i^2}\right]\, \left[\det_{1\leq j,k\leq m_l} L_{k}^{\alpha-1/2}(r_j^2) \right] \ , 
\end{eqnarray}
where $Z_{m_l,\alpha}$ is the normalization constant. The determinant of Laguerre polynomials can be expressed in terms of a Vandermonde determinant giving
\begin{eqnarray}\label{vdm}
\left[\det_{1\leq j,k\leq m_l} L_{k}^{\alpha-1/2}(r_j^2) \right]  \propto \prod_{1\leq j<k\leq m_l} (r_k^2 - r_j^2) \;.
\end{eqnarray}   
The joint PDF of the positions of the $m_l$ Fermions in the ground state is given by
\begin{eqnarray}\label{jpdf_heff}
p_{\alpha}(r_1,r_2,\cdots, r_{m_l})= \left|\Psi_\alpha(r_1,r_2,\cdots, r_{m_l}) \right |^2 = A_{m_l, \alpha}\left[\prod_{i=1}^{m_l} r_i^{2\alpha} \, e^{-r_i^2}\right] \, \prod_{1\leq j,k\leq m_l}(r_k^2-r_j^2)^2 \;,
\end{eqnarray} 
 where $A_{m_l, \alpha}$ is a normalization constant, such that $\int dr_1 \cdots \int dr_{m_l} \, p_{\alpha}(r_1,r_2,\cdots r_{m_l})=1$. 
 
In this model, the CDF of the position of the rightmost fermion can be written as
\begin{eqnarray}\label{prod_heff}
P_{\alpha,m_l}(w) = {\rm Prob.}[\max\{r_1, r_2, \cdots, r_{m_l} \} \leq w] = \Big\langle \prod_{i=1}^{m_l} (1 - I_w(r_i)) \Big \rangle
\end{eqnarray} 
where $I_w(r_i) = 1$ if $r_i \geq w$ and $0$ otherwise. The average $\langle \cdots \rangle$ is carried out  with respect to the joint distribution $p_\alpha(r_1, \cdots, r_{m_l})$ in Eq. (\ref{jpdf_heff}). Following similar steps as used from Eqs. (\ref{prod}) to (\ref{cdf_det_2}), but specializing now to the $1d$ case, we get
\begin{eqnarray}\label{eq:identity}
P_{\alpha,m_l}(w) = \det_{1 \leq i,j \leq m_l}\left[ \delta_{i,j} - \int_{w}^\infty \chi_{n_i,l}(r) \chi_{n_j,l}(r) \, dr\right] \equiv P_l(w,m_l) \;,
\end{eqnarray} 
with $\alpha = l + (d-1)/2$ and $m_l = N_l(\mu)+1$. Thus the quantity $P_l(w,m_l)$ that appears in Eq. (\ref{product_det}) can be interpreted as the CDF of the position of the rightmost fermion in the ground state of this effective $1d$ quantum model.

Making a change of variable $\lambda_i = r_i^2$, the joint PDF of the $\lambda_i$'s variables can be written from $p_\alpha(r_1, r_2, \cdots, r_{m_l})$ in Eq. (\ref{jpdf_heff}) as
\begin{equation}
\rho_{\alpha}(\lambda_1,\lambda_2,\cdots, \lambda_{m_l}) ={1\over Z'_{m_l,l}}  
 \left[\prod_{i=1}^{m_l}  \lambda_i^{\alpha-1/2} e^{-\lambda_i}\right]\prod_{1\leq j<k\leq m_l}(\lambda_k-\lambda_j)^2\label{secl}
\end{equation}
where $Z'_{m_l,l} $ is another normalization constant and we recall that $\alpha = l+(d-1)/2$ and $m_l = N_l(\mu)+1$. This is exactly the joint PDF of the eigenvalues of the Laguerre-Wishart random matrices with Dyson index $\beta = 2$ \cite{meh91,for10}. 

Let us quickly recall the Laguerre-Wishart ensemble of random matrices. A Wishart matrix is an $m \times m$ product matrix of the form $W = X^\dagger X$ where $X$ is an $M\times m$ rectangular random matrix and $X^\dagger$ is its Hermitian conjugate, which is of size $m \times M$ \cite{wis28}. If one takes the distribution of the entries of $X$ to be Gaussian, drawn from the joint PDF 
\begin{equation}\label{jpdf_X}
P(X)\propto \exp\left[-\frac{\beta}{2}{\rm Tr}(X^\dagger X)\right],
\end{equation}
where the Dyson index $\beta=1$ for real matrices and $\beta=2$ for 
complex matrices. Without any loss of generality, we consider here $M \geq  m$ (the case $M\leq m$ can be analyzed by exchanging $M$ and $m$ \cite{VMO07}). The product matrix $W$ of size $m \times m$ has $m$ non-negative real eigenvalues $\lambda_1, \cdots, \lambda_m$ whose joint PDF is given by 
\begin{equation}
P_{\rm LW}(\lambda_1,\cdots,\lambda_m) \propto  
\exp\left(-\frac{\beta}{2}\sum_{i=1}^{m}\lambda_i\right) \left[\prod_{i=1}^{m}  \lambda_i^{\frac{\beta}{2}(M-m+1) - 1}\right]\prod_{1\leq j,k\leq m}|\lambda_k-\lambda_j|^\beta,\label{wishart}
\end{equation}
where the subscript LW refers to the Laguerre-Wishart ensemble. Even though in the original Wishart matrices $M$ and $m$ are integers, the more general Laguerre-Wishart ensemble corresponds to Eq. (\ref{wishart}) where the parameter $M-m > -1$ can be any real number.  
We thus see that the distribution in Eq. (\ref{secl}) is the same as that of the eigenvalues of the $\beta =2$ Laguerre-Wishart ensemble
upon the identifications: 
\bea\label{def_mM}
m=m_l=N_l(\mu)+1 \;\;\;\; {\rm and} \;\;\;\; M =M_l=m_l+ l+\frac{d}{2} -1 \;,
\eea
where $N_l(\mu)$ is given in Eq. (\ref{Nl}). Going back to Eq. (\ref{eq:identity}) and recalling that $\lambda_i=r_i^2$, we see that $P_l(w,m)$ can then be interpreted as the CDF of the maximal eigenvalue in the Laguerre-Wishart ensemble 
\begin{eqnarray}\label{lambdamax}
P_l(w,m_l) = {\rm Prob.} \, \left[\max \left\{\lambda_1, \lambda_2, \cdots, \lambda_{m_l}\right\} \leq w^2 \right] \;.
\end{eqnarray}
For later convenience, we define a random variable in each $l$-sector
\bea\label{rmax_l}
r_{\max,l} = \sqrt{\max \left\{\lambda_1, \lambda_2, \cdots, \lambda_{m_l}\right\}} \;.
\eea
Clearly, it follows from Eqs. (\ref{lambdamax}) and (\ref{rmax_l}) that
\bea\label{rmax_l_2}
P_l(w,m_l) = {\rm Prob.}(r_{\max,l} \leq w) \;.
\eea
Therefore from Eq. (\ref{product_det}), we see that the CDF of $r_{\max}$, $P(w,N)$, is the product of CDFs of $(\mu-d/2)$ (interpreted as the integer part) independent random variables $r_{\max,l}$, one for each $l$. Each of these random variables is the square root of the maximum eigenvalue of a Laguerre-Wishart matrix with size parameters $m_l = N_l(\mu)+1$ and $M_l =m_l+ l+\frac{d}{2} -1$. The Laguerre-Wishart ensembles labelled by $l$ are independent of each other. Thus $r_{\max}$ corresponds to the extreme of a set of independent but non-identically distributed random variables. 

It is important to note that the above statements and formula are correct {\it for arbitrary finite $N$} provided
the last energy level is fully occupied, so that the ground state is unique.
In the next subsection we study the asymptotics of $P(w,N)$ for large $N$, by analyzing the product structure in Eq. (\ref{product_det}) and we will show that the same Gumbel law as in Eqs. (\ref{fred.7}) and (\ref{amu}) will emerge out of this analysis.    

{\bf Remark for $d=1$:} We note that in $d=1$, using the one to one correspondence between the fermion positions and the eigenvalues of the $N \times N$ GUE matrices, it follows that 
\begin{eqnarray}\label{abs_GUE}
P(w,N) = {\rm Prob.}[\max\{|x_1|, \ldots, |x_N|\} \leq w] \;,
\end{eqnarray}
where $x_i$'s denote the fermion positions (or equivalently the GUE eigenvalues). Going back to Eq.~(\ref{product_det}) and noting that, in $d=1$, there are only two allowed values of $l$, namely $l=0$ and $l=1$ with degeneracies $g_1(0)=g_1(1)=1$, it follows that the product in Eq. (\ref{product_det}) has only two terms
\begin{eqnarray}\label{prod_d1}
P(w,N) = P_0(w,m_0) \, P_1(w,m_1) 
\end{eqnarray}
with $\mu = N-1/2$, $m_0 = 1+{\rm int}((N-1)/2)$, $m_1 = 1+{\rm int}((N-2)/2)$. Thus for any $N$ there is a factorization into two sectors, which corresponds to two independent squared Laguerre ensembles [clarifying Eq. (\ref{rmax_l})]. This is reminiscent of the factorization of the joint PDF of the absolute values of the eigenvalues of GUE into two squared Laguerre ensembles (sometimes called chiral Laguerre ensembles) \cite{For06, Ede15, Bor15}.  Again, these statements are valid for any finite $N$.

\subsection{Large $N$ analysis of $P(w,N)$ in Eq. (\ref{product_det})}
\label{sec:product}

%In this subsection, we analyse the limiting form of $P(w,N)$ for large $N$, from Eq. (\ref{product_det}). It turns out that the $d=1$ case is very different %from the $d \geq 2$ case.  

Our starting point is the expression for $P(w,N)$ in terms of a product, given in Eq. (\ref{product_det}), where $P_l(w,m_l)$ is given
in Eq. (\ref{lambdamax}), which corresponds to the CDF of a Laguerre-Wishart matrices with size parameters $m_l = N_l(\mu)+1$ and $M_l = m_l+ l+\frac{d}{2} -1$. In the following, in order to lighten the notations, we will drop the subscript $l$ from both $m_l$ and $M_l$, and denote them simply by $m$ and $M$.

Since the Fermi energy $\mu \propto N^{1/d}$ [see  Eq.~(\ref{mu_4})], the limit of large $N$ corresponds to large $\mu$. Consequently, from Eq.~(\ref{def_m}), $m_l\equiv m$ is large for large $N$ and hence
we need to analyze $P_l(w,m)$ for large $m$. The distribution of the largest eigenvalue $\lambda_{\max}$ of Laguerre-Wishart matrices in the large $m$ limit has been well studied in the literature \cite{joh00,johnstone}. Let us recall the main results that we will need here for our analysis of $P_l(w,m)$. 

Consider a complex ($\beta=2$) Laguerre-Wishart ensemble with size parameters $m, M$ such that both $m$ and $M$ are large with their ratio
held fixed. In this limit, the largest eigenvalue $\lambda_{\max}$ approaches to an identity in law
\begin{eqnarray}\label{TW2_1}
\lambda_{\max} = A_{m,M} + \sigma_{m,M} \, \chi_2 \;,
\end{eqnarray}
where
\begin{eqnarray}\label{A_sigma}
A_{m,M} = \left( \sqrt{m} + \sqrt{M}\right)^2 \;\;\;\; {\rm and} \;\;\;\; \sigma_{m,M} = \left( \sqrt{m} + \sqrt{M}\right) \left(\frac{1}{\sqrt{m}} + \frac{1}{\sqrt{M}} \right)^{1/3}
\end{eqnarray}
and $\chi_2$ is a random variable, independent of $m$ and $M$, whose CDF is given by the Tracy-Widom distribution 
\begin{eqnarray}
{\rm Prob.} \left(\chi_2 \leq s \right) = {\cal F}_2(s) \;.
\eea 
This function ${\cal F}_2(s)$ can be expressed in terms of a special solution of a Painlev\'e II equation and has the large $s \to \infty$ asymptotic tail
\begin{equation}\label{large_s}
{\cal F}_2(s) \sim 1- \frac{1}{16\pi s^{\frac{3}{2}}}\exp\left(-\frac{4}{3}\,s^{\frac{3}{2}}\right) \;\;\;\;  {\rm as }\;\;\;\; s\to \infty \;.
\end{equation}

To adapt these results to our problem, we proceed as follows. Let us first express $A_{m,M}$ and $\sigma_{m,M}$ in terms of $\mu$ and $l$ for large $\mu$. From Eq. (\ref{def_m}) for large $\mu$, we can ignore the integer part and write 
\begin{eqnarray}\label{m_large}
m \simeq \frac{1}{2} \left(\mu - l -\frac{d}{2} + 2 \right) \simeq \frac{1}{2} \left(\mu - l \right) \;,
\end{eqnarray} 
anticipating that $\mu - l$ will be large (to be verified a posteriori). Similarly, $M$ in Eq. (\ref{def_mM}) is given, for large $\mu$, by
\bea\label{M_large}
M \simeq \frac{1}{2} (\mu+l + \frac{d}{2}) \simeq \frac{1}{2}(\mu + l) \;.
\eea
The subsequent analysis is different for $d=1$ and $d \geq 2$ and we will analyse the two cases separately. 

\vspace*{0.5cm}
\noindent
{\bf Case $d=1$}. In this case, there are only two allowed values of $l$, $l=0$ and $l=1$. Hence from Eqs. (\ref{m_large}) and (\ref{M_large}), it follows that, for large $\mu$, $m \simeq M \simeq \mu/2$ for both $l=0$ and $l=1$. Consequently from Eq. (\ref{A_sigma}), one gets
\begin{eqnarray}\label{A_d1}
A_{m,M} \simeq 2\,\mu \;\;\; {\rm and} \;\;\; \sigma_{m,M} \simeq 2 \mu^{1/3} \;.
\eea  
Using $r_{\max,l} = \sqrt{\lambda_{\max}}$ from Eq. (\ref{rmax_l}), it follows from Eq. (\ref{TW2_1}) that, both for $l=0$ and $l=1$ 
\bea\label{rmax_d1}
r_{\max,l} =  \left( A_{m,M} + \sigma_{m,M} \,\chi_2\right)^{1/2} \simeq \sqrt{2 \mu} + \frac{1}{\sqrt{2}} \mu^{-1/6}\, \chi_2 \;,
\eea
where we have expanded the square root for large $\mu$ and kept the two leading order terms. For $l=0$ and $l=1$, the CDF of $r_{\max,l}$ thus has  the scaling form
\bea\label{cdf_d1}
P_l(w,m) = {\rm Prob.}(r_{\max,l} \leq w) = {\cal F}_2\left(\sqrt{2} \,\mu^{1/6}(w-\sqrt{2\,\mu})\right) \;.
\eea
Using this result in Eq. (\ref{prod_d1}), it follows that in $d=1$, the cumulative distribution of $r_{\max}$ is the square of the Tracy-Widom distribution
\bea\label{final_d1}
P(w,N) \simeq \left[ {\cal F}_2\left(\sqrt{2} \,\mu^{1/6}(w-\sqrt{2\,\mu})\right)\right]^2 \;,
\eea 
in the large $N$ limit. This result has a natural interpretation which follows from Eq. (\ref{abs_GUE}). Indeed,  Eq. (\ref{abs_GUE}) can be re-expressed in terms of the joint distribution of $x_{\min} = \min\{x_1, \cdots, x_N\}$ and $x_{\max} = \max\{x_1, \cdots, x_N\}$ as
\begin{eqnarray} \label{min_max}
P(w,N) = {\rm Prob.}(x_{\max} \leq w, x_{\min} \geq -w) \;.
\end{eqnarray}
While, for finite $N$, $x_{\max}$ and $x_{\min}$ are strongly correlated, they become independent for large $N$, leading to a decoupling of Eq. (\ref{min_max}) into a product 
\begin{eqnarray} \label{min_max_2}
P(w,N) \simeq {\rm Prob.}(x_{\max} \leq w) \, {\rm Prob.}(x_{\min} \geq -w) \;.
\end{eqnarray}
Since each of these probabilities is given, for large $N$, by the Tracy-Widom limiting distribution, one gets the
square of the Tracy-Widom CDF in Eq. (\ref{final_d1}). Note that, even in the large $N$ limit, this factorization does not
hold for all $w$. It holds only when both the maximum and the minimum are close to their respective typical values, i.e. when $|w - \sqrt{2\mu}| \sim {\cal O}(\mu^{-1/6})$. In contrast when $|w - \sqrt{2\mu}| \sim {\cal O}(\sqrt{\mu})$, i.e., when both the maximum and the minimum are close to the center of the trap, this factorization no longer  holds anymore. Instead $P(w,N)$ is described by a large deviation form that was computed exactly in \cite{dea08}.

\vspace*{0.5cm}
\noindent
{\bf Case $d \geq 2$}. In this case, the possible values of $l$ run from 0 to $\mu$, for large $\mu$. 
Substituting the large $\mu$ expansions given in Eqs. (\ref{m_large}) and (\ref{M_large}) in Eq. (\ref{A_sigma}), expanding for large $\mu$ and for the moment keeping $l/\mu = \ell$ fixed, we obtain
\begin{eqnarray}\label{A_sigma_large_mu}
A_{m,M} = \mu(1 + \sqrt{1 - \ell^2}) \;\;\; {\rm and} \;\;\; \sigma_{m,M} = (2\,\mu)^{1/3} \frac{(1+\sqrt{1-\ell^2})^{2/3}}{(1-\ell^2)^{1/6}} \;\;\; {\rm where} \;\;\; \ell = \frac{l}{\mu} \;.
\eea 
Using $r_{\max,l} = \sqrt{\lambda_{\max}}$ from Eq. (\ref{rmax_l}), it follows from Eq. (\ref{TW2_1}) that   
\bea\label{rmax_expand}
r_{\max,l} = \left( A_{m,M} + \sigma_{m,M} \,\chi_2\right)^{1/2} \simeq \sqrt{A_{m,M}} + \frac{1}{2} \frac{\sigma_{m,M}}{\sqrt{A_{m,M}}} \, \chi_2 \;,
\eea
where we have expanded the square root for large $\mu$ and kept the two leading order terms. This then gives
\bea\label{rmax_expand2}
r_{\max,l} \simeq C(\ell) \, \sqrt{\mu} + D(\ell)\, \mu^{-1/6} \, \chi_2 \;,
\eea
where
\bea\label{def_cd}
C(\ell) = \left( 1 + \sqrt{1-\ell^2} \right)^{1/2}  \;\;\; {\rm and}  \;\;\; D(\ell) =  2^{-2/3}\,\left[\frac{1+\sqrt{1-\ell^2}}{1-\ell^2} \right]^{1/6}
\eea
where we recall that $\ell= l/\mu$. The CDF of $r_{\max,l}$ has thus the scaling form
 \bea\label{P_l_final_1}
 P_l(w,m) = {\rm Prob.}(r_{\max,l} \leq w) = {\cal F}_2\left(s_\ell \right) \; \quad , \quad s_{\ell}=\frac{w-C(\ell)\,\sqrt{\mu}}{D(\ell) \, \mu^{-1/6}} \;,
 \eea
in the region of the typical fluctuations, where $s_{\ell}={\cal O}(1)$ and we recall that ${\cal F}_2(s)$ is the CDF of the GUE TW distribution. We substitute this result (\ref{P_l_final_1}) into the product in Eq. (\ref{product_det}) and obtain
\bea\label{P_w_F2}
P(w,N) = \exp{\left[S(w)\right]} \;\;\; {\rm where}  \;\;\; S(w) = \sum_{l\geq 0} g_d(l) \ln {\cal F}_2\left(\frac{w-C(\ell)\,\sqrt{\mu}}{D(\ell) \, \mu^{-1/6}} \right) \;.
\eea
For this formula to hold we need that the minimum value $s_{\min} = \min_{\ell \in [0,1]} s_{\ell}$, of the argument $s_{\ell}$ of the TW distribution is in the range $s_{\min} = {\cal O}(1)$ (see discussion below).

For $d \geq 2$, the sum over $l$ in Eq. (\ref{P_w_F2}) runs up to $\mu$ where $\mu$ is large. Hence the asymptotic large $N$ analysis of $P(w,N)$ in this case is rather different from the $d=1$ case discussed above. It turns out that the leading contribution to the sum comes from the large $l$ regime (where $l = {\cal O}(\mu^{2/3})$ for large $\mu$, a fact to be verified a posteriori), where we can use the asymptotic form of $g_d(l)$ for large $l$ in Eq. (\ref{gd}). Furthermore, recalling that $l = \ell\, \mu$, it is clear that in the large $\mu$ limit, one can replace the sum over $l$ by an integral over $\ell$. This gives  
\bea\label{Sw_asympt}
S(w) \simeq  \frac{2 \, \mu^{d-1}}{\Gamma(d-1)} \int_0^1 d\ell\, \ell^{d-2} \, \ln {\cal F}_2\left(\frac{w-C(\ell)\,\sqrt{\mu}}{D(\ell)\, \mu^{-1/6}} \right)
\eea 
where $C(\ell)$ and $D(\ell)$ are given in Eq. (\ref{def_cd}). 

We must now analyze the argument $s_{\ell}=\mu^{2/3} (\hat w - C(\ell))/D(\ell)$ of the Tracy-Widom CDF,
where we have defined $\hat w=w/\sqrt{\mu}$. Since
$-\ln {\cal F}_2(s_{\ell})$ is a positive decreasing function, the integral will be a priori dominated by the 
minimum value of $s_{\ell}$, i.e. $s_{\min}$, as $\ell$ varies. 
The analysis depends on the value of $\hat w$, which we
suppose fixed as $\mu$ becomes large.
There are only two possible local minima for $s_{\ell}$
on the interval $\ell \in [0,1]$, one at $\ell=0$, where locally
$s \simeq (2\mu)^{2/3}( 2^{-1/6}(\hat w- \sqrt{2}) - \frac1{2^{1/6} 8} (\hat w-2 \sqrt{2}) \ell^2)$
and one at $\ell=1$ where $s \simeq (2\mu)^{2/3} (\hat w - 1)(1-\ell^2)^{1/6}$. For $\hat w<1$ there is
only the minimum at $\ell=0$. For $1 \leq \hat w \leq 2 \sqrt{2}$ both minima exist, and the absolute minimum
is at $\ell=0$ for $\hat w<\sqrt{2}$, while it is at $\ell=1$ for $\sqrt{2} < \hat w < 2 \sqrt{2}$. 

We now focus on the regime near the edge of the Fermi gas, $\hat w=w/\sqrt{\mu}=\sqrt{2}$, which
turns out to contain the typical fluctuation regime (this can be anticipated from the results of section \ref{sec:limit}). We will see that the leading contribution comes from the small $\ell$ regime (of order ${\cal O}(\mu^{-1/3})$). Let us examine the contribution of the minimum of $s_{\ell}$ at $\ell=0$ (from the discussion of the previous paragraph). 
One can write the argument of ${\cal F}_2$ in Eq. (\ref{Sw_asympt}) as
%One can further verify that the leading contribution comes from the small $z$ regime (of order ${\cal O}(\mu^{-1/3})$ as we will see now) where the argument of ${\cal F}_2$ in Eq. (\ref{Sw_asympt}) reads, to leading order
\bea\label{cd_small}
s_{\ell}=\frac{w-C(\ell)\,\sqrt{\mu}}{D(\ell) \, \mu^{-1/6}} \simeq \tilde w + \frac{1}{4} \,\ell^2 \,\mu^{2/3} + {\cal O}(\ell^4)   \;\;\; {\rm where} \;\;\; \tilde w = (w - \sqrt{2 \mu})\sqrt{2} \,\mu^{1/6} \;.
\eea
Clearly one must rescale $\ell = 2 v \mu^{-1/3}$ in the integral over $\ell$ and one obtains in the 
large $\mu$ limit
\bea
S(w) \simeq \frac{2^d \, \mu^{\frac{2(d-1)}{3}}}{\Gamma(d-1)} 
\int_0^{+\infty} dv\, v^{d-2} \, \ln {\cal F}_2(\tilde w + v^2) \label{Sw22} 
\eea
This result is correct as long as $\hat w - \sqrt{2}$ vanishes at large $\mu$, i.e. 
$|\tilde w| \ll \mu^{2/3}$. For other values of $\hat w$ one needs to use,
instead of the typical fluctuation formula given in Eq. (\ref{P_l_final_1}), a large deviation formula,
which goes beyond the scope of this paper. 

Let us now study the regime of typical fluctuations. Denoting, as in Eqs. \eqref{ab.1}--\eqref{ab.2}, 
$\tilde w = a_\mu + z/(2\,\sqrt{a_\mu})$ and anticipating that $a_\mu$ is large and $z={\cal O}(1)$, we can now use
the asymptotic tail of ${\cal F}_2(s)$ for large $s$ in Eq. (\ref{large_s}), 
inserting $s = a_\mu + z/(2\,\sqrt{a_\mu})   +  \,v^2$ in \eqref{Sw22}, one finds for large $\mu$
\bea
S(w) \simeq -\frac{2^d \, \mu^{\frac{2(d-1)}{3}}}{16 \pi \Gamma(d-1) a_\mu^{3/2}} 
\int_0^{+\infty} dv\, v^{d-2} \,
e^{- \frac{4}{3} a_\mu^{3/2} - 2 \sqrt{a_\mu} v^2 - z} \label{Sw2} \;.
\eea
We can now substitute, as in \eqref{ab.1}-\eqref{ab.2}, $\tilde w = a_\mu + z/(2\,\sqrt{a_\mu})$
where $a_\mu$ is large and $z={\cal O}(1)$. Expanding, the leading behavior at large $\mu$ of the expression (\ref{Sw2}) leads to evaluate the
integral 
\bea
\int_0^{+\infty} dv\, v^{d-2} \, e^{- 2 \sqrt{a_\mu} v^2} = 2^{-\frac{d+1}{2}} a_\mu^{\frac{1-d}{4}} \Gamma\left(\frac{d-1}{2}\right) \;, 
\eea
leading exactly to 
\bea
S(w) \simeq - B_d\,\, e^{- \frac{4}{3}\, a_\mu^{3/2}+ 
\frac{2(d-1)}{3}\, \ln \mu - \frac{(d+5)}{4}\, \ln (a_\mu) - 2\, 
\sqrt{a_{\mu}}\, b_{\mu}\, z} \;,
\eea
where $B_d$ was given in \eqref{Bd}.
Hence, we arrive exactly at
the result in Eq. (\ref{fred.5}), and \eqref{fred.7} in section II B, i.e. a Gumbel distribution for $r_{\max}$, although by a quite different method. The present calculation
further shows that the angular momentum which contributes to the
value of $r_{\max}$ are of order $l = \ell \mu \sim \mu^{2/3}$ (see the interpretation below).

Let us comment on the contribution of the second minimum of $s_{\ell}$ at $\ell=1$, 
which corresponds to high angular momentum states, and turns out to be negligible for $\hat w=\sqrt{2}$.
A naive analysis of the continuum integral (\ref{Sw_asympt}) would suggest an additional contribution
to $S(w)$ corresponding to values of $\ell =1 - {\cal O}(\mu^{-4})$. 
However in the original sum (\ref{P_w_F2}), $l$ is an integer
which runs up to $l_{\max}={\rm int}(\mu - d/2)$. Hence the continuum integral
must be cutoff at values of $\ell =1 - {\cal O}(\mu^{-1})$, which corresponds to
$s_{\min} \sim \sqrt{\mu}$, hence leading to an exponentially small 
contribution to $S(w)$, i.e. $\sim {\cal O}(\mu^{d-2} e^{- \mu^{3/4}}) \ll \mu^{\frac{2(d-1)}{3}}$. In fact one can directly analyze the contribution coming from  $l=l_{\max}$, corresponding to states with $n=0$
and having a degeneracy $\sim \mu^{d-2}$ (see Fig. \ref{Fig_levels}). The associated radial wavefunction 
$\chi_{0,l_{\max}}(r)$ given by (\ref{expr_chi}) can be shown to be peaked 
around $r^* \simeq \sqrt{\mu} \ll$ which is far from the trap center but is also far from the edge of the Fermi gas   (recall $r_{\rm edge}=\sqrt{2 \mu}$). One can calculate
explicitly for this state, using formula \eqref{P_l} (which is  a simple one-dimensional
determinant as $m_l=1$) the probability
\bea
P_{l_{\max}}(w,m_{l_{\max}}=1) = 1-  \frac{\int_{w}^{+\infty} dr ~ r^{2 l_{\max}+ d-1} e^{-r^2} }{\int_{0}^{+\infty} dr ~ r^{2 l_{\max}+ d-1} e^{-r^2}} \simeq 1 - {\cal O}(e^{- 2 \mu}) \;,
\eea 
leading to an exponentially small 
contribution to $S(w)$, i.e. $\sim {\cal O}(\mu^{d-2} e^{- 2 \mu}) \ll \mu^{\frac{2(d-1)}{3}}$.
Hence there is no  additional contribution due to  these maximal angular momentum states, in agreement with 
the analysis above.

\section{Extensions to other observables and finite temperature} 
In this section we give some further probabilistic interpretations 
as well as new results. We study counting statistics near the edge, 
extreme value statistics in non-spherical domains,
the distribution of $x_{\max}$, and present some extensions to finite temperature.
 
\subsection{Poisson statistics near the expected maximal radial distance} 

Let us define the number of fermions $N_w$ outside the sphere centered at the
origin and of radius $w$ in the ground state. From the determinantal structure 
of the quantum probability of $N$ fermions at zero temperature (\ref{eq:Rn_det}), we know that the
Laplace transform of the distribution of $N_w$ can be written as a Fredholm determinant
\bea
\langle e^{- p N_w} \rangle_0 = {\rm Det}( I - (1- e^{-p}) I_w K_\mu) \;.
\eea 
We will restrict our analysis to values of $w$ near the edge $w \approx r_{\rm edge}$
where, if $p={\cal O}(1)$ and in $d>1$, we can use the single trace approximation as
in Section \ref{sec:limit}. In addition we consider the region around the average
value of $r_{\max}$, hence we set
\bea
w = \sqrt{2 \mu} + \frac{1}{2 \mu^{1/6}} \left(a_\mu + \frac{z}{\sqrt{2 a_\mu}} \right) \;,
\eea 
where $z$ is a parameter of order unity. Following the same steps as in that Section we
easily arrive at 
\bea
\langle e^{- p N_w} \rangle_0 =  \exp \left( (e^{-p} - 1) e^{-z} \right) \;.
\eea 
This means that $N_w$ has a Poissonian statistics with parameter $e^{-z}$, i.e.
\bea
{\rm Prob}(N_w = k) = \frac{e^{- k z}}{k!} \exp(- e^{-z} ) \;,
\eea 
where the random variable is $N_w$ and $z$ a parameter. Since one has 
$\langle N_w \rangle_0=e^{-z}$ the Poisson distribution crosses over to a
Gaussian distribution for $z < 0$ and $|z| \gg 1$. This is the case in the region
close to $r_{\rm edge}$, where $\tilde w = {\cal O}(1) \ll a_\mu$.

\subsection{Other domains} 
\label{sec:otherdomains} 

Studying systems in higher dimensions opens up the possibility to study a wider variety of extreme value problems than in one dimension. For instance one can ask what is the distribution of the farthest fermion from the origin when one looks within a certain solid angle, or what is the distribution of the maximal coordinate in a given direction.
Here we address these issues.

Let us come back to the single trace approximation in Eq. \eqref{fred.2} for the probability that $r_{\max} <w$, namely
\bea
{\rm Prob.}(r_{\max} \leq w) \simeq \exp\left[-\int_{{\bf x} \in {\cal D}} K_\mu({\bf x},{\bf x})\, d{\bf x}\right]
\simeq \exp{\left[ - S_d \left(\frac{r_{\rm edge}}{w_N}\right)^{d-1} 
\int_{\tilde w}^{\infty} F_d(s) \, ds \right]}  \;,
\eea 
where here the domain ${\cal D}$ is the region $|{\bf x}|> w$, and we have chosen a $w$ such that 
$w-r_{\rm edge} = {\cal O}(w_N)$. One can interpret the second equality, roughly, as saying that there are
a number of order ${\cal N}_{\rm eff} \sim S_d (r_{\rm edge}/w_N)^{d-1}$ 
independent regions contributing to the maximal radius (naturally leading then to a Gumbel
distribution). Said differently, it means that along the surface of the Fermi droplet, i.e. in the angular
direction, the positions of the fermions within the domain ${\cal D}$,
are correlated over distances of order $w_N$. This is only a rough interpretation, a more
detailed and precise one was presented in the previous section, but it is sufficient
for our purpose here.

It is then reasonable to assume that the same approximation should hold for more general domains, 
as long as the number of independent regions ${\cal N}_{\rm eff}$ is large as $N \to +\infty$,
leading again to a Gumbel distribution (this condition also ensures that the higher power traces in the expansion of the Fredholm determinant (\ref{Fredholm_id}) remain subdominant).
Let us for instance consider a domain ${\cal D}$ defined as the region $|{\bf x}|> w({\boldsymbol \theta})$ 
where ${\boldsymbol \theta}$ is an angular direction (i.e. a $(d-1)$-dimensional angular vector, e.g. 
a simple angle $\theta \in [0,2 \pi[$ for $d=2$). 
Here $w({\boldsymbol \theta})$ is an arbitrary curve, but to be able to use the edge scaling form of the kernel
we will choose it such as $w({\boldsymbol \theta}) \geq r_{\rm edge} + {\cal O}(w_N)$ and we define 
\bea
\tilde w({\boldsymbol \theta})= \frac{w({\boldsymbol \theta}) -r_{\rm edge}}{w_N}= (w({\boldsymbol \theta})-\sqrt{2\mu})\, \sqrt{2}\, \mu^{1/6}\, \label{wtilde2} \;.
\eea
We then ask what is the probability that there are no fermions in the domain ${\cal D}$, i.e.
the ``hole probability'' associated to the domain ${\cal D}$. 
One example in $d=2$ is $\tilde w(\theta)= \tilde w$ in the angular sector $\theta \in [0,\theta_m]$,
and else $\tilde w(\theta)=+\infty$, and this hole probability then becomes the probability that $r_{\max}\leq w$ 
{\it restricted to the fermions in the angular sector $[0,\theta_m]$} (see Fig.~\ref{Fig_thetam}). In general it can
be expressed as a Fredholm determinant as in 
\eqref{prod_0}, \eqref{genf}, with $I_w({\bf x})$ the indicator function of the domain ${\cal D}$. 
Keeping again the leading single trace term we find (defining $N_{\cal D}$ the number of fermions in the domain~${\cal D}$)
\bea\label{hole}
{\rm Prob.}(N_{\cal D} =0)  \simeq \exp{\left[ - \left(\frac{r_{\rm edge}}{w_N}\right)^{d-1} \int d{\boldsymbol \theta}
\int_{\tilde w({\boldsymbol \theta})}^{\infty} F_d(s) \, ds \right]} \;,
\eea 
and similar manipulations as in Section \ref{sec:limit} lead to a Gumbel distribution for a suitably scaled
$\tilde w$ (with different constants $a_\mu$ and $b_\mu$). 

\begin{figure}
\includegraphics[width = 0.4 \linewidth]{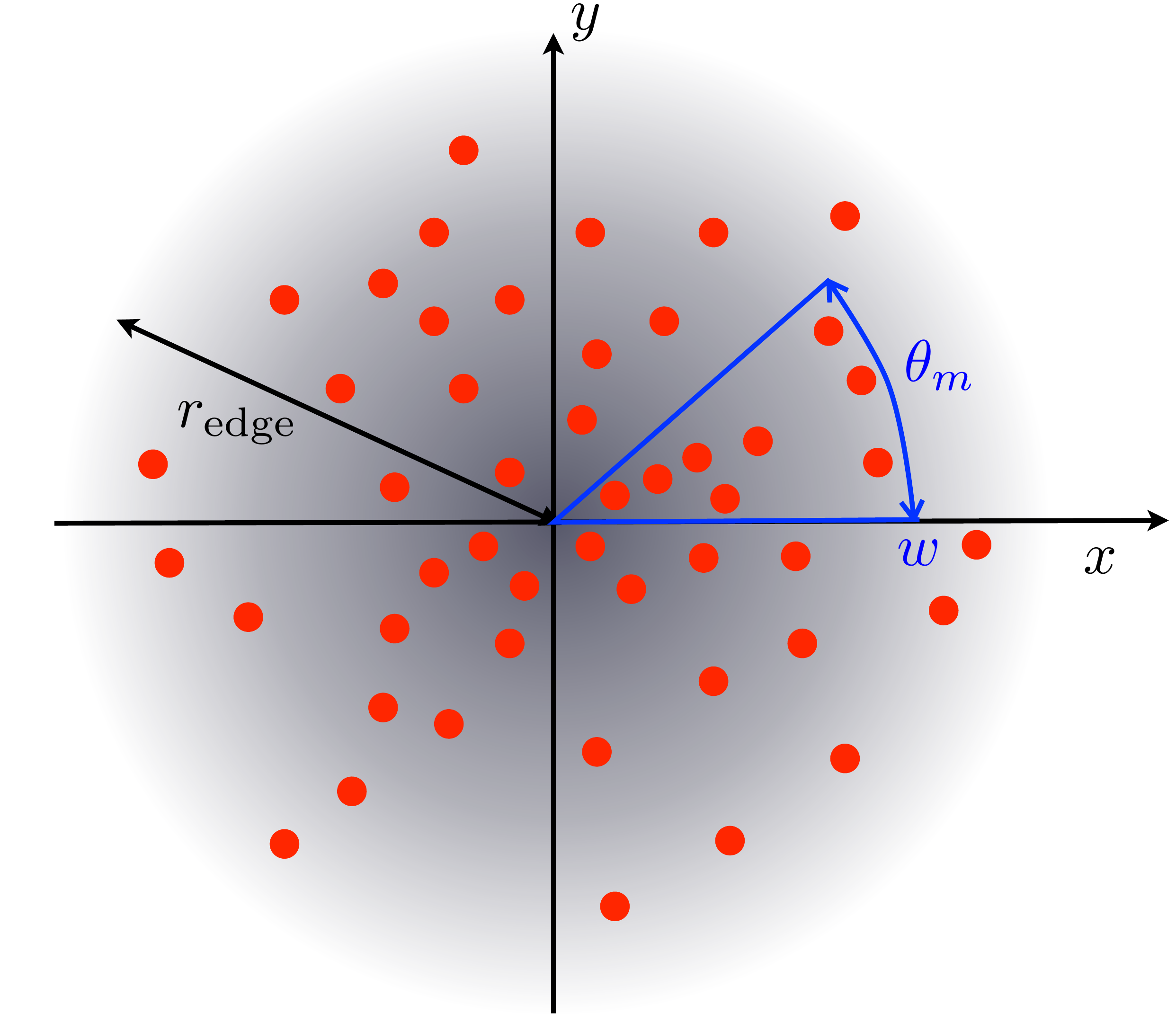}
\caption{Illustration of the domain ${\cal D}$, in $d=2$, corresponding to $w(\theta)= w$ in the angular sector $\theta \in [0,\theta_m]$, and elsewhere $\tilde w(\theta)=+\infty$. The hole probability associated with ${\cal D}$ is the probability that $r_{\max}\leq w$ 
{\it restricted to the fermions in the angular sector $[0,\theta_m]$}.}\label{Fig_thetam}
\end{figure}

Let us give an example of application in $d=2$ (another example being given in the next subsection). First consider the probability, denoted
$P_{\theta_m}(w,N)$, that $r_{\max} \leq w$, restricted to the fermions in the angular sector $[0,\theta_m]$.
The effective number of independent regions is ${\cal N}_{\rm eff} \sim \theta_m \, \mu^{2/3}$. Hence 
if the angular sector is sufficiently large, i.e. if $\theta_m \gg \mu^{-2/3} \sim N^{-1/3}$ one obtains
a Gumbel distribution,
$P_{\theta_m}(w,N) \simeq \exp( - e^{- z})$ where $\tilde w=a_\mu + \frac{1}{2 \sqrt{a_mu}} z$, 
with a parameter $a_\mu$ solution of
\begin{equation}
\frac{2^{-7/2}}{\sqrt{\pi}} \theta_m\, e^{- \frac{4}{3}\, a_\mu^{3/2}+
\frac{2}{3}\, \ln \mu - \frac{7}{4}\, \ln (a_\mu)}=1 \;,
\label{ab.10}
\end{equation}
leading to $a_\mu \simeq \left[\frac{1}{2} \ln (\mu \, \theta_m^{3/2})\right]^{\frac{2}{3}}$.
If the angular sector becomes too small, {\it i.e.}, $\theta_m$ of order, or smaller than
$\mu^{-2/3} \sim N^{-1/3}$ we expect a crossover to a Tracy-Widom regime,
which would be quite interesting to study, but is beyond the scope of the present 
paper. Note that the scale $\theta_m \sim \mu^{-2/3}$ is consistent with the
angular momentum scale $l \sim \theta_m^{-1} \sim \mu^{2/3}$ which was
shown in the previous Section to control the value of $r_{\rm max}$. 

Note finally that the arguments in this section have been mostly qualitative, and to make them rigorous
(and study the crossover to the $d=1$ limit) 
would require a general analysis of the higher traces for a larger class of domains ${\cal D}$ which goes
beyond the scope of the present paper.

\subsection{Probability distribution of $x_{\max}$ in $d=2$} 
\label{sec:xm} 

Another interesting question is to calculate the PDF of the first coordinate, i.e. its abscissa denoted by $x_{\max}$, of
the rightmost fermion in $d=2$ dimensions. 
We start with a first method, using results of the previous subsection.
This corresponds to the choice $w(\theta)=w/\cos(\theta)$. One can
check a-posteriori that one can restrict the analysis to small angles $\theta$, hence we write 
$w(\theta)= w (1 + \frac{\theta^2}{2})$.  Thus we obtain
\bea
{\rm Prob.}(x_{\max} \leq w)  \simeq \exp{\left[ - 2 \mu^{2/3} \int d \theta
\int_{\tilde w + \mu^{2/3} \theta^2}^{\infty} F_d(s) \, ds \right]} \;,
\eea 
where we have used the definition \eqref{wtilde2}. 
Expanding as $\tilde w = a_\mu + \frac{1}{2 \sqrt{a_\mu}} z$ we obtain again a Gumbel
distribution for the variable $z$, i.e. ${\rm Prob.}(x_{\max} \leq w) = e^{-e^{-z}}$, with
now the following condition to determine $a_\mu$
\begin{equation}
\tilde B_2(\mu) \; e^{- \frac{4}{3}\, a_\mu^{3/2}+
\frac{2}{3}\, \ln \mu - \frac{7}{4}\, \ln (a_\mu)}=1  \quad , \quad \tilde B_2(\mu) = \frac{1}{(8 \pi)^{3/2}}  
\int_{-\infty}^\infty d\theta e^{- 2 a_\mu^{1/2} \mu^{2/3} \theta^2} = \frac{1}{32 \pi} a_\mu^{-1/4} \mu^{-1/3} \;,
\label{ab.11}
\end{equation}
which leads to $- \frac{4}{3}\, a_\mu^{3/2} + \frac{1}{3} \ln \mu - 2 \ln a_\mu - \ln(32 \pi)=0$. Hence to leading order one has
\bea
a_\mu \simeq \left( \frac{1}{4} \ln \mu \right)^{2/3}   \;. \label{amuxmax} 
\eea
Note that the angles involved in this maximization are $\theta \sim \theta_\mu = a_\mu^{-1/4} \mu^{-1/3}$ 
which justifies the expansion of the $\cos (\theta)$ for small $\theta$ (i.e., its approximation by a parabola). The maximum
$x_{\max}$ is thus attained for fermions which deviate at most of order $\theta_\mu$ from the axis
(up to powers of $\ln \mu$). Also the number of independent regions is
${\cal N}_{\rm eff} = \mu^{1/3} a_\mu^{-1/4} \gg 1$ as $N$ becomes large,
which a posteriori justifies the calculation. To summarize, 
we find
\begin{equation}
x_{\max}= \sqrt{2\mu} + \frac{1}{\mu^{1/6}}\, \left[\frac{1}{\sqrt{2}}\, a_\mu
+ \frac{1}{2\sqrt{2\, a_\mu}}\, Z_G\right]\,,
\label{fred.77}
\end{equation}
where ${\rm Prob.}[Z_G\le z]= \exp[-e^{-z}]$ and $a_\mu$ is given by \eqref{amuxmax}. Comparing with
the similar formula for $r_{\max}$, in Eq.~\eqref{fred.7}, \eqref{amu} we see that 
$x_{\max}$ is smaller than $r_{\max}$ but its fluctuations are larger. 

We will now show that this result can be obtained from an exact calculation, similar to the one
of Section \ref{sec:angular}, which we present here only in $d=2$ for simplicity -- the generalization to
higher $d$ being straightforward.
To study $x_{\max}$ it is more convenient to use the eigenbasis of the single particle problem
\bea
\phi_{n_x,n_y}(x,y) = \phi_{n_x}(x) \phi_{n_y}(y) \quad , \quad \phi_{k}(x) = \left(\frac{1}{\sqrt{\pi} 2^k k!}\right)^{1/2} H_k(x) e^{-x^2/2} \;,
\eea 
where $\phi_k(x)$ are the eigenstates of the $1d$ harmonic oscillator, and $H_k$ the Hermite polynomial of degree $k$. As in formula (\ref{cdf_det_2}) we have 
\begin{eqnarray}\label{cdf_det_2new}
{\rm Prob.}(x_{\max} \leq w) &=&  \det_{1 \leq i,j \leq N} \left[\delta_{ij} - \int_{x \geq w} dx \, dy  \,\phi_{n_{x,i}, n_{y,i}}(x,y) \phi_{n_{x,j},
n_{y,j}}(x,y)   \right] \nonumber \\
&=& \det_{1 \leq i,j \leq N} \left[\delta_{ij} - \delta_{n_{y,i},n_{y,j}} \int_{x \geq w} dx  \,\phi_{n_{x,i}}(x) \phi_{n_{x,j}}(x)   \right]  
\end{eqnarray} 
where the $(n_{x,i},n_{y,i})$, $i=1, \ldots, N$ denote the occupied states (note that in Eq. (\ref{cdf_det_2new}) we have used that the eigenfunctions $\phi_k(x)$ are real). In the $N$-body ground state all 
single particle states with energy $\epsilon(n_x,n_y) = n_x + n_y + 1 \leq \mu$ are 
occupied, leading to the relation between $N$ and the Fermi energy $\mu$: $N=\frac{1}{2} \mu (\mu+1)$
(we consider $\mu$ to be equal to the energy of the last occupied state, hence it is an integer). 
\begin{figure}
\includegraphics[width = 0.4\linewidth]{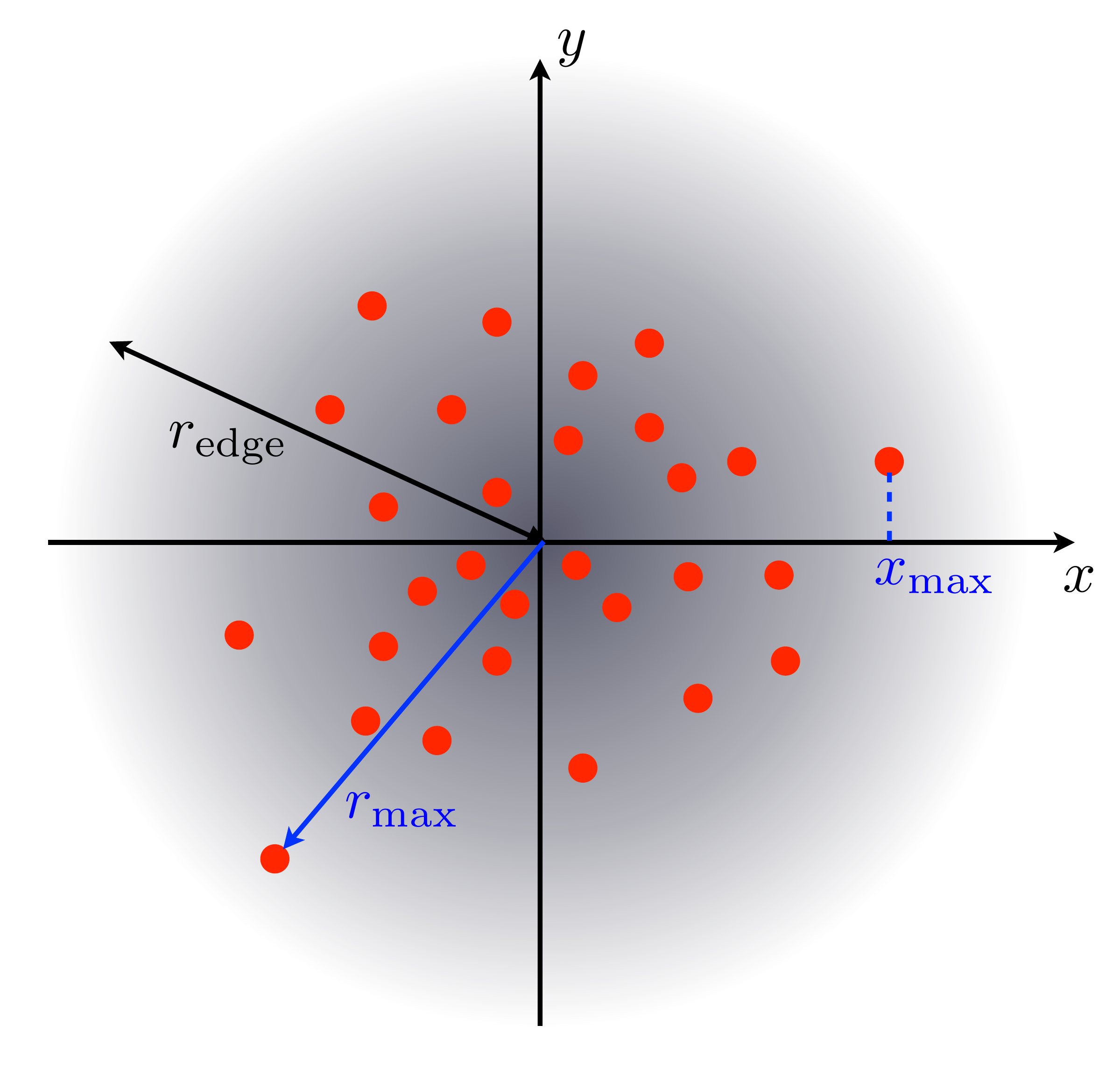}
\caption{Illustration of the position $x_{\max}$ of the rightmost fermion for a two-dimensional system confined in a harmonic trap, as studied in [see Eq. (\ref{product_detnew})]. In the large $N$ limit, the PDF of $x_{\max}$ properly centered and scaled converges also to a Gumbel distribution.} \label{fig_xmax}
\end{figure}

As in the previous calculation (see section \ref{sec:angular}) this determinant (\ref{cdf_det_2new}) has a block structure and hence one
can write 
\begin{eqnarray}
{\rm Prob.}(x_{\max} \leq w) &=&
\label{product_detnew}
\prod_{n_y \geq 0}^{\mu-1} \, \Big[P_{n_y}(w,\mu-1-n_y) \Big]
\end{eqnarray} 
where here 
\bea
P_{n_y}(w,N_{n_y}) = \det_{1 \leq i,j \leq N_{n_y}} \left[\delta_{ij} -
\int_{x \geq w} dx  \,\phi_{n_{x,i}}(x) \phi_{n_{x,j}}(x)   \right]  
\eea 
is the cumulative distribution function (CDF) of the position of the right-most fermion among $N_{n_y}$ fermions in
a one-dimensional harmonic potential. This CDF has been studied in Ref.~\cite{dea15a} -- as well as in the RMT literature~\cite{for10} -- and is known to take
the form of a Tracy-Widom distribution at large $N_y$, i.e. one has
\bea
x_{\max,n_y} = \sqrt{2 (\mu-1-n_y)} + \frac{1}{\sqrt{2}} (\mu-1-n_y)^{-1/6} \chi_2 \;,
\eea
where the CDF of $\chi_2$ is ${\cal F}_2(s)$ (to be compared with formula \eqref{rmax_expand}
for the study of $r_{\max}$). 
Setting $n_y=\ell \mu$ in the limit of large $\mu$
and following similar steps as in Section \ref{sec:angular}, one finds
${\rm Prob}(x_{\max} \leq w) \simeq \exp( S(w))$ where $S(w)$ can be written as
a sum over the states 
\bea
S(w) = \mu \int_0^1 d \ell  \ln {\cal F}_2 \left( (w - \sqrt{2 \mu(1-\ell)}) \sqrt{2} \mu^{1/6} (1-\ell)^{1/6} \right) \;,
\eea
where we have made the  approximation $\sum_{n_y=0}^{\mu-1} \simeq \mu \int_0^1 d\ell$. 
Inserting the right tail of the Tracy-Widom distribution, we find again that the 
region which dominates the integral is for small $\ell$. Hence, expanding in $\ell$ close to $\ell=0$, the argument of ${\cal F}_2$ can be replaced by $\tilde w + \ell \mu^{2/3}$. A similar calculation to that  in Section \ref{sec:angular}
then allows us  to recover exactly the result of Eq.~\eqref{fred.77}, with exactly the same condition determining 
$a_\mu$, leading to the estimate in \eqref{amuxmax}.

\subsection{Higher traces and the regime $\tilde w = {\cal O}(1)$} 

It is also interesting to ask what is the effect of the higher order traces in \eqref{hole_proba}. As we 
have hinted at above they only contribute to subdominant corrections  in the  determination of  the distribution of the {\rm typical} fluctuations of $r_{\max}$ [see Eq. (\ref{fred.7})], when $\tilde w \sim a_\mu \gg 1$. However
these terms become important if one studies the regime $\tilde w={\cal O}(1)$, as we now discuss.

In Ref. \cite{dea15a,dea16} we have shown that for ${\bf x},{\bf y}$ near a point $r_{\rm edge} {\bf n}$, where ${\bf n}$ is a unit vector
(on the unit sphere), the kernel takes the scaling form
\bea \label{kernelTVedge0} 
 && K_{\mu}({\bf x},{\bf y})  \simeq \frac{1}{w_N^d} {\cal K}^{{\rm edge}}_{d, {\bf n}}
 \left(\frac{ {\bf x} - r_{\rm edge} {\bf n} }{w_N}, 
 \frac{ {\bf y} - r_{\rm edge} {\bf n} }{w_N}\right) 
 \eea
 % \bea \label{scaling_kernelTVedge0} 
% && {\cal K}^{{\rm edge}}_{d,b}({\bf a}, {\bf b}) = 2^{2\over 3}\int_{-\infty}^\infty {du\over \exp(-bu)+1}
% \int \frac{d{\bf q}}{(2 \pi)^d}  e^{-i {\bf q} \cdot ({\bf a} - {\bf b})  } {\rm Ai}\left(2^{\frac{2}{3}} q^2 + \frac{a_n+b_n+2u}{2^{1/3}}\right) \;,
% \eea
where the $d$-dimensional edge kernel can be written as
 \bea
 {\cal K}^{{\rm edge}}_{d,{\bf n}}({\bf a}, {\bf b}) &=& 
% \int_{-\infty}^\infty {du\over \exp(-bu)+1}
% \int \frac{d{\bf q}_t}{(2 \pi)^{d-1}}  e^{-i {\bf q}_t \cdot ({\bf a}_t - {\bf b}_t)  } {\rm Ai}\left(
% a_n +{\bf q}_t^2 + u \right){\rm Ai}\left(
% b_n +{\bf q}_t^2 + u \right),  \label{kernelTVedge} \\
% & =&  
 \int \frac{d{\bf q}_t}{(2 \pi)^{d-1}}  e^{-i {\bf q}_t \cdot ({\bf a}_t - {\bf b}_t)  }  
K_{\rm Ai}(a_n +{\bf q}_t^2,b_n +{\bf q}_t^2)\;, \label{edgeKd2} 
 \eea
in terms of the standard Airy kernel, $K_{\rm Ai}(a,b) = \int_0^{+\infty} du \, {\rm Ai}(a+u) {\rm Ai}(b+u)$. In Eq. (\ref{edgeKd2}) and below we
decompose any vector ${\bf a}={\bf a}_t + a_n  {\bf n}$ into a normal (scalar) component $a_n = {\bf a}\cdot {\bf n}$ and a tangential component ${\bf a}_t$ (a $(d-1)$-dimensional vector). We have indicated
by an additional subscript that the edge kernel explicitly depends on ${\bf n}$. Let us note the following
interesting property 
\bea
\int d{\bf b}_t \, {\cal K}^{{\rm edge}}_{d,{\bf n}}({\bf a}, {\bf b}) \, {\cal K}^{{\rm edge}}_{d,{\bf n}}({\bf b}, {\bf c}) 
=  \int \frac{d{\bf q}_t}{(2 \pi)^{d-1}}  e^{-i {\bf q}_t \cdot ({\bf a}_t - {\bf c}_t)  }  
K_{\rm Ai}(a_n +{\bf q}_t^2,b_n +{\bf q}_t^2) K_{\rm Ai}(b_n +{\bf q}_t^2,c_n +{\bf q}_t^2) \label{product1} 
\eea 
which allows (by iteration, see below and Appendix \ref{sec:higher}) to calculate any power of the kernel in terms of Airy functions. 

Using this scaling form \eqref{kernelTVedge0} one can reduce the problem of calculating the trace in Eq.~(\ref{Trn}) to
free integrations over ${\bf a}_2, \cdots, {\bf a}_n$ together with an integral over ${\bf n}$ (over the unit sphere)
and over $a_{1,n}$ the radial part of ${\bf a}_1$. Hence it is easy to see that (see Appendix \ref{sec:higher} for details) 
\bea
P(w,N) &\simeq& \exp\left( -  \left(\frac{r_{\rm edge}}{w_N}\right)^{d-1} 
\sum_{p=1}^\infty \frac{1}{p} \int d{\bf n} {\rm Tr}\left[ \delta_{{\bf a}_{1,t}, {\bf 0}} 
(\tilde I_{\tilde w,{\bf n}}\,  {\cal K}^{{\rm edge}}_{d,{\bf n}})^p\right]  \right) \label{traces1} \\
&=& \exp\left(   \left(\frac{r_{\rm edge}}{w_N}\right)^{d-1} 
 \int d{\bf n}  \ln {\rm Det}(I - \delta_{{\bf a}_{1,t}, {\bf 0}} 
\tilde I_{\tilde w,{\bf n}} {\cal K}^{{\rm edge}}_{d,{\bf n}}  ) \right) \label{det1} \;,
\eea 
where $\int d{\bf n}$ denotes the integral over the unit sphere, 
$\delta_{{\bf a}_{1,t}, {\bf 0}}$ denotes the projector on the subspace ${\bf a}_{1,t}= {\bf 0}$,
and the trace is over all vectors ${\bf a}_1, \cdots, {\bf a}_n$. In the last line
the traces have been summed into a Fredholm determinant. If one studies $r_{\max}$, then the 
operator 
$\tilde I_{\tilde w,{\bf n}}({\bf a})$ is the projector on $a_n \geq \tilde w$, but the above expression is
valid for more general domains (such as studied in Section \ref{sec:otherdomains}). 
In the case of the calculation of $r_{\max}$, using rotational invariance it is easy to see
(see Appendix \ref{sec:higher}) that all the traces in (\ref{traces1}) are independent of ${\bf n}$
hence one can replace $\int d{\bf n}$ by $S_d$ and pick one particular fixed direction ${\bf n}$.
The calculation of the traces then uses an iteration of (\ref{product1}) 
\bea
 {\rm Tr}\left[ \delta_{{\bf a}_{1,t}, {\bf 0}} 
(\tilde I_{\tilde w,{\bf n}}\,  {\cal K}^{{\rm edge}}_{d,{\bf n}})^p\right]  = 
\int \frac{d{\bf q}_t}{(2 \pi)^{d-1}}  Tr[ (P_{\tilde w + {\bf q}_t^2} K_{\rm Ai})^p] \;,
\eea 
where $P_s$ is the standard notation for the projector on $[s,+\infty[$. Summing up the traces
we obtain the formula for the probability that $r_{\max} \leq w$:
\bea
P(w,N) &\simeq& \exp\left(  S_d \left(\frac{r_{\rm edge}}{w_N}\right)^{d-1} 
\int \frac{d{\bf q}_t}{(2 \pi)^{d-1}}  \ln {\rm Det}(I - P_{\tilde w + {\bf q}_t^2} K_{\rm Ai}) \right) \\
&=&  \exp\left(  S_d \left(\frac{r_{\rm edge}}{w_N}\right)^{d-1} 
\int \frac{d{\bf q}_t}{(2 \pi)^{d-1}} \ln {\cal F}_2(\tilde w + {\bf q}_t^2) \right) \label{largedev} \;,
\eea 
where ${\cal F}_2(s)= {\rm Det}(I - P_{s} K_{\rm Ai})$ is the GUE Tracy-Widom CDF. 
This expression is valid for $N$ large, and for $\tilde w = {\cal O}(1)$, as well as in the 
regime $\tilde w \sim a_\mu \gg 1$ studied previously. Hence this expression shows the
large deviation form for the PDF of $r_{\max}$, when $\tilde w = {\cal O}(1)$, which
is an intermediate left deviation regime for the present problem (note that there exists another
left large deviation regime, deeper inside the Fermi cloud, whose study is left for the future).
The exponent in (\ref{largedev}) is the associated rate function. Here we have 
obtained it exactly in terms of the GUE-TW function ${\cal F}_2(s)$, in the case 
of a spherically symmetric domain.

Performing the angular integral in \eqref{largedev}, and denoting $|{\bf q}_t|=v$
it is immediate to see, using the identities $S_d S_{d-1}/(\pi)^{d-1}=2^d/\Gamma(d-1)$
and $r_{\rm edge}/w_N=2 \mu^{2/3}$,
that the formula \eqref{largedev} is identical to Eq. (\ref{Sw22}), obtained 
in Section \ref{sec:product} by quite different methods. There it was studied only
in the typical fluctuation regime $\tilde w \simeq a_\mu \gg 1$, but the
present analysis shows that these formula holds down to the regime $\tilde w = {\cal O}(1)$. 
%Of course, in the end it is an equivalent form to the exact formula 
%\eqref{product_det} in the large deviation regime. {\red Pierre says: I suggest one does
%a more careful job in section Laguerre and shows exact identity of the 2 formula} 

The explicit calculation of the rate function for more general domains in
this regime $\tilde w = {\cal O}(1)$, i.e. of the reduced Fredholm determinant 
for non-spherically symmetric domains, remains challenging. The analysis of
the previous section shows, that a simplification should also arise
in %the case of the domain associated to 
the calculation of 
$x_{\max}$. Finally, in the region $\tilde w \sim a_\mu$ this Fredholm determinant can be approximated 
by the leading single trace, and one recovers the typical fluctuations studied in
this paper (see Appendix \ref{sec:higher} for further details).

\subsection{Finite temperature} 

We now briefly sketch the effect of a finite temperature on the distribution of the maximal radius, denoted by $r_{\max}(T)$. As above, we restrict ourselves  to the case
of the harmonic oscillator. As was shown in \cite{dea16} 
the corresponding hole probability can be calculated at finite temperature. For this purpose, one uses the equivalence of the
canonical (fixed $N$) and grand-canonical (fixed chemical potential $\tilde \mu$) ensembles 
in the large $N$ limit, and the fact that in the grand-canonical ensemble the JPDF of the positions of the
fermions form a determinantal process \cite{dea16,Joh07}. As a result the formula \eqref{hole_proba} is
still valid but $K_\mu$ is replaced by the finite temperature kernel, described in \cite{dea16}.
Since here we study $r_{\max}(T)$, i.e., the edge of the Fermi gas, we can
again use the scaling form of the finite temperature kernel near the edge, which is valid
in the regime where the temperature $T \sim N^{1/(3 d)}$: this is the regime 
on which we focus from now on. More precisely one
defines the dimensionless parameter \cite{dea15b,dea16}
\bea
b = (\Gamma(1+d) N)^{1/(3 d)}/T  \;.
\eea 
We now use the same (single trace) approximation as we used at $T=0$ (whose validity is discussed below) 
and we obtain, following similar steps to those in Section \ref{sec:limit} 
\bea\label{P_1}
P(w,N) \simeq \exp{\left[ -S_d\, \left(\frac{r_{\rm edge}}{w_N}\right)^{d-1} \int_{\tilde w}^{\infty} F^{\rm edge}_{d,b}(z) \, dz \right]} \;.
\eea
As before, we have written again $\tilde w= \frac{w-r_{\rm edge}}{w_N}= (w-\sqrt{2\mu})\, \sqrt{2}\, 
\mu^{1/6}$ where here and below $\mu$ is the Fermi energy, i.e. the $T=0$ chemical potential, 
as defined in Section \ref{sec:GS} [see Eq. (\ref{expr_mu})].
This formula involves the finite temperature
version of the scaling function of the density, obtained in \cite{dea16} as
\bea
F^{{\rm edge}}_{d,b}(z)  = \frac{2^{2-d} \pi^{\frac{1-d}{2}}}{\Gamma(\frac{d-1}{2})} 
\int_0^{+\infty} dq \, q^{d-2} \, F^{{\rm edge}}_{1,b}(z + q^2) \;, \label{link} 
\eea 
where 
 \bea
F_{1,b}^{ \rm edge}(s) = \int_{-\infty}^{+\infty} du \frac{{\rm Ai}(s+u)^2}{1 + e^{-b u}} \;,
\eea 
is the $d=1$ finite temperature density scaling function. Clearly we again need the large positive $\tilde w$ asymptotics of Eq.~(\ref{P_1}), which requires the large argument asymptotics of
the function $F_{d,b}(z)$ (right tail), which was not given in \cite{dea16}. Let us recall that the right tail of $F_{1,b}(z)$ was obtained in \cite{LargeDev_KPZ,dea16} and shown to depend on the
parameter $\tilde s= s/b^2$ as follows
\bea\label{F1_right_tail}
F^{\rm edge}_{1,b}(s) \simeq \begin{cases}
& \dfrac{1}{4 b^2 \sqrt{\tilde s} \sin(2 \pi \sqrt{\tilde s}) } \exp \left(- \frac{4}{3} s^{3/2} \right)  
\quad , \quad 1 \ll s <  \frac{b^2}{4} \\
\\
& \dfrac{1}{\sqrt{4 \pi b} } \exp \left(- b s + \frac{b^3}{12} \right) \quad , \quad s > \frac{b^2}{4} \;.
\end{cases} 
\eea 
Hence it exhibits a transition between a stretched exponential tail 
to a pure exponential decay for $s > s_c=b^2/4$. Thus, 
for a fixed value of the reduced temperature $b$ (not necessarily large), 
the decay is always exponential. The pre-exponential factor
further exhibits a crossover from the two limiting cases indicated above, in the
vicinity of $\tilde s=1/4$ as given in \cite{LargeDev_KPZ}. Note that the first
regime in Eq.~\eqref{F1_right_tail} for $\tilde s \to 0$, when inserted in \eqref{link}, gives back the
$T=0$ behavior \eqref{Fdz.2}.

Let us start with the ``thermal'' regime where we can use the second form in \eqref{F1_right_tail},
i.e. $s> \frac{b^2}{4}$. Inserting into \eqref{link} we find the right tail 
\bea
F^{{\rm edge}}_{d,b}(s) \simeq \dfrac{1}{(4 \pi b)^{d/2}} \exp \left(- b s + \frac{b^3}{12} \right) \;,
\eea 
%\bea
%\int_{\tilde w}^{+\infty} F^{{\rm edge}}_{d,b}(s) \simeq \dfrac{1}{(4 \pi)^{d/2} b^{1+ \frac{d}{2}}} \exp \left(- b \tilde w + \frac{b^3}{12} \right)
%\eea 
which gives
\bea
P(w,N) \simeq \exp \left[ -S_d\, \left(\frac{r_{\rm edge}}{w_N}\right)^{d-1} 
\dfrac{1}{(4 \pi)^{d/2} b^{1+ \frac{d}{2}}} e^{ - b \tilde w + \frac{b^3}{12} } \right] \;,
\eea
where we recall that $\frac{r_{\rm edge}}{w_N}= 2 \mu^{2/3} = 2 (\Gamma(d+1) N)^{2/(3 d)}$. 
Hence we obtain again a Gumbel distribution in terms of a variable $z$ now defined as
\bea
P(w,N) \simeq e^{- e^{-z}}   \quad , \quad \tilde w = a_\mu + \frac{1}{b} z
\eea 
where $a_\mu$ is here determined by the condition
\bea \label{amu_b}
A_d \; e^{ - b a_\mu + \frac{2}{3}(d-1) \ln \mu + \frac{b^3}{12} -  (1+ \frac{d}{2}) \ln b }  = 1 \;,
\eea 
with $A_d=\dfrac{S_d 2^{d-1}}{(4 \pi)^{d/2}}=1/\Gamma(d/2)$. This condition (\ref{amu_b}) yields the leading order behavior
\bea\label{expr_amub}
a_{\mu} = a_{\mu,b} \simeq \frac{2}{3 b}(d-1) \ln \mu + \frac{b^2}{12} - \frac{d+2}{2 b} \ln b  + \frac{1}{b} \ln A_d \;,
\eea
which is our final result for the distribution of $r_{\max}(T)$ in the thermal regime,
i.e., of fixed $b$ and large $N$ (i.e. large $\mu$)
\begin{equation}
r_{\max}(T)= \sqrt{2 \mu} + \frac{1}{\sqrt{2} \mu^{1/6}}\,  \left( a_{\mu,b}
+ \frac{1}{b}\, Z_G \right)
\label{fred.7_b}
\end{equation}
where ${\rm Prob.}[Z_G\le z]= \exp[-e^{-z}]$. We note that this result remains valid for $d=1$ 
in the limit $b \ll 1$, and exactly matches with the result obtained in \cite{dea16} (discarding 
the term $b^2/12$ which is clearly subdominant in this limit). To check this we used the fact -- see \eqref{min_max_2} -- that in
$d=1$, the PDF of $r_{\max}(T)$ is given by the maximum of two independent Gumbel random variables, each
obtained in Eq. (156) of \cite{dea16}.

Clearly, comparing this result in Eqs.  (\ref{expr_amub}) and (\ref{fred.7_b}) with the result obtained at $T=0$ and given in Eqs. \eqref{fred.7} and (\ref{amu}), we see that there is a crossover
at very low temperature $b = b^*$ from the thermal regime to the $T=0$ regime, with
\bea\label{eq:b*}
b^* \sim \frac{4}{3} \left(\frac{d-1}{2} \ln \mu\right)^{1/3} \;.
\eea 
This crossover occurs between two Gumbel distributions with different parameters. 
The study of the details of this crossover is left for the future.

Let us also indicate the result for the distribution of $x_{\max}(T)$ 
at finite temperature in $d=2$. By a similar calculation, using similar arguments as in Section \ref{sec:xm}, we find
\begin{equation}
x_{\max}(T)= \sqrt{2 \mu} + \frac{1}{\sqrt{2} \mu^{1/6}}\,  \left( a_{\mu,b}
+ \frac{1}{b}\, Z_G \right)
\label{fred.7_b_xmax}
\end{equation}
where ${\rm Prob.}[Z_G\le z]= \exp[-e^{-z}]$ is a Gumbel variable,
with
\bea\label{expr_amub_xmx}
a_{\mu} = a_{\mu,b} \simeq \frac{1}{3 b} \ln \mu + \frac{b^2}{12} - \frac{5}{2 b} \ln b  - \frac{1}{b} \ln (2 \sqrt{\pi}) \;.
\eea
Note that we find that the amplitude of the fluctuations of $x_{\max}(T)$ and $r_{\max}(T)$  are identical
but that on average $r_{\max}(T) > x_{\max}(T)$, as expected. 

Finally, recalling from \cite{dea16} (section VII-D) that the formula \eqref{edgeKd2} 
extends to any value of $b$, one easily sees that the expression in (\ref{largedev}) 
extends to finite temperature, at fixed $b$, as follows
\bea
P(w,N) &\simeq& \exp\left(  S_d \left(\frac{r_{\rm edge}}{w_N}\right)^{d-1} 
\int \frac{d{\bf q}_t}{(2 \pi)^{d-1}} \ln {\cal F}_{2,b}(\tilde w + {\bf q}_t^2) \right) \label{largedev_T} \;,
\eea 
where here we denote ${\cal F}_{2,b}(s)= {\rm Det}(I - P_{s} K^{\rm edge}_{b})$ the finite
temperature generalization of the GUE Tracy-Widom CDF, where $K^{\rm edge}_{b}$
is given in section
V-C of \cite{dea16}.

\section{Discussion}

In this paper, we have investigated analytically the distribution of the maximal radial distance
from the trap center of $N$ non-interacting fermions in a $d$-dimensional harmonic trap at $T=0$. 
We have shown that, in the large $N$ limit, the cumulative distribution, appropriately centered and scaled, 
converges to the Gumbel distribution for $d>1$. In $d=1$, we have shown that the limiting cumulative distribution is given by the
square of the Tracy-Widom distribution for GUE. If one thinks of the dimension $d$ as a continuous 
parameter, clearly the distribution crosses over from the squared Tracy-Widom to the Gumbel form. It would
be interesting to investigate the precise crossover function that interpolates between these two distinct forms, as $d$ crosses the critical value $d=1$. 

How universal is the large $N$ scaling form of the cumulative distribution with respect to the shape of the trapping potential? We have shown that the limiting form of the cumulative distribution can be derived solely from the asymptotic
properties of the edge kernel. However it has been recently shown that the edge kernel, properly centered and scaled, 
is universal, i.e., independent of the details of the trapping potential, for a broad class of confining potentials. 
Consequently, the limiting distribution of the maximal radial distance is also universal, with respect to this class of trapping potentials. Note that a different extreme value observable is the smallest distance to the center of
the trap, $r_{\min}$. Its distribution is related to the probability to have a hole of radius $r_{\min}$ at the
center, a question which has been studied in \cite{Torquato} , but for which more remains to be done in $d>1$.

Another natural question is how this limiting distribution depends on temperature $T\geq 0$. 
Previous studies of matrix models related to finite temperature fermions \cite{LaguerreT,2DfiniteT} 
have not addressed questions about extreme value statistics (see however \cite{Joh07} in connection with $1d$ fermions). 
In $d=1$, the distribution of the position of the rightmost fermion in a harmonic trap has recently been studied 
for finite temperatures \cite{dea15a,dea16}. The cumulative distribution, properly centered and scaled, was found to converge to a Gumbel form (i) at very high temperature $T \gg N$ and (ii) also at intermediate temperature where $N^{1/3} \ll T \ll N$. The scale factors associated with these Gumbel forms depend on $T$ and $N$, however these dependencies are rather different in the two regimes. In this
work we have investigated what happens in higher dimensions as temperature increases. 
We have studied only the regime $T \sim N^{1/(3 d)}$ and shown that the distribution of $r_{\max}(T)$ is also
a Gumbel distribution, although with different parameters to those at $T=0$. The other extreme limit is when $T \to \infty$, where we expect that the fermions behave as independent classical particles in a harmonic potential (the Pauli exclusion principle becomes completely irrelevant). Consequently, the maximal radial distance will correspond to the maximum among a set of $N$ independent and identically distributed random variables, each with a Gaussian tail. Hence we would expect that the limiting cumulative distribution of the maximal distance, properly centered and scaled, will again converge to the standard Gumbel form in all dimensions $d > 1$. How the limiting distribution crosses over from the zero temperature Gumbel form to the infinite temperature Gumbel form as $T$ increases, is an interesting open problem.

Finally, let us stress that some of the predictions of the present theory are in principle testable in
cold atom experiments, due to the recent progress in Fermi gas quantum gas 
microscopes 
\cite{Cheuk:2015,Haller:2015,Parsons:2015,Omran2015,Toronto,Bakr2009,Sherson2010,Bakr2010}. 
Even though in general the atoms are interacting, the current experimental setups also allow one to tune the strength of the interactions to zero and are able to probe the non-interacting limit \cite{BDZ08,GPS08}.
In such microscopes, individual atoms from a correlated many-body system can be imaged in situ with a resolution comparable to the inter-particle spacing, providing direct access to the modulus of the system's wave-function.  A quantum gas microscope gives access not only to the real space distribution but, using a time of flight technique, can also provide the momentum distribution of the dilute gas. Hence both
$r_{\max}$, $p_{\max}$ and $x_{\max}$ can be studied. In practice one needs to include the 
thermal effects, which are not negligible in the experiments. We hope that the present
study will stimulate efforts in this direction.

\appendix
\section{Higher order terms in the trace expansion}\label{sec:higher}

In subsection \ref{sec:limit} we used the first term in the trace expansion to derive the distribution of the farthest fermion. Here we consider the behavior of the higher order terms and show that they are indeed negligible
in the regime of typical fluctuations. To begin with, we analyze the second term in the trace expansion which is proportional to
\begin{equation}
{\rm Tr}( [I_w K_\mu]^2) =\iint_{\cal D} d{\bf x}\,d{\bf y} K_\mu({\bf x},{\bf y})^2.
\end{equation}
where the integration domain ${\cal D}$ corresponds to $|{\bf x}|,\ |{\bf y}| > w$.
In general we can write 
\begin{equation}
\int_{\cal D} d{\bf y} K_\mu({\bf x},{\bf y})^2 = \int_{\cal D} d{\bf y} K_\mu(r{\bf n},{\bf y})^2, \label{k2}
\end{equation}
where $r=|{\bf x}|$ and ${\bf n}={\bf x}/r$. Clearly isotropy implies that Eq. (\ref{k2}) must be independent of 
${\bf n}$. We can thus write
\begin{equation}
{\rm Tr}( [I_wK]^2) =  S_d \int_w^\infty dr \, r^{d-1}\int_{\cal D} d{\bf y} K_\mu(r{\bf n},{\bf y})^2,
\end{equation}
where ${\bf n}$ is an arbitrary direction. We now write $r= r_{\rm edge} +a'_n$ and, anticipating
that the integral over ${\bf y}$ is dominated by points close to ${\bf x}$, we also write
${\bf y} = r_{\rm edge}{\bf n} + {\bf b}'= (r_{\rm edge} +  b'_n){\bf n} + {\bf b}'_t$
where ${\bf b}'_t$ is the transverse component of the vector ${\bf y}$ in the plane whose normal is ${\bf n}$.
Using the fact that $r_{\rm edge}$ is large means that we can write
%\begin{equation}
%{\rm Tr}( [I_wK]^2) =  S_d r^{d-1}_{\rm edge}\int_{w-r_{\rm edge}}^\infty da'_n \int_{\cal D} d{\bf x}'K([r_{\rm edge} + a'_n]{\bf n},{\bf x}')^2
%\end{equation}
%We now use the fact that the integral over ${\bf x}'$ is dominated by points close to ${\bf x}$ and thus write
%\begin{equation}
%{\bf x}' = r_{\rm edge}{\bf n} + {\bf b}'= (r_{\rm edge} +  b'_n){\bf n} + {\bf b}'_t,
%\end{equation}
% This then gives
\begin{equation}\label{eq_app1}
{\rm Tr}( [I_wK]^2) \simeq  S_d r^{d-1}_{\rm edge}\int_{w-r_{\rm edge}}^\infty da'_n \int_{w-r_{\rm edge}}^\infty db'_n\int d{\bf b}_t' K([r_{\rm edge} + a'_n]{\bf n}, [r_{\rm edge} + b'_n]{\bf n} + {\bf b}_t')^2.
\end{equation}

In Ref. \cite{dea15a,dea16} it was shown that for ${\bf x},{\bf y}$ near a point $r_{\rm edge} {\bf n}$, where ${\bf n}$ is a unit vector
(on the unit sphere), the kernel takes the scaling form
\bea \label{kernelTVedge0_app} 
 && K_{\mu}({\bf x},{\bf y})  \simeq \frac{1}{w_N^d} {\cal K}_{d,{\bf n}}^{{\rm edge}}
 \left(\frac{ {\bf x} - r_{\rm edge} {\bf n} }{w_N}, 
 \frac{ {\bf y} - r_{\rm edge} {\bf n} }{w_N}\right) \;, 
 \eea
 % \bea \label{scaling_kernelTVedge0} 
% && {\cal K}^{{\rm edge}}_{d,b}({\bf a}, {\bf b}) = 2^{2\over 3}\int_{-\infty}^\infty {du\over \exp(-bu)+1}
% \int \frac{d{\bf q}}{(2 \pi)^d}  e^{-i {\bf q} \cdot ({\bf a} - {\bf b})  } {\rm Ai}\left(2^{\frac{2}{3}} q^2 + \frac{a_n+b_n+2u}{2^{1/3}}\right) \;,
% \eea
where  ${\cal K}_{d,{\bf n}}^{{\rm edge}}({\bf a}, {\bf b})$ is given in \eqref{edgeKd2} 
% \bea
% {\cal K}^{{\rm edge}}({\bf a}, {\bf b}) &=& 
%% \int_{-\infty}^\infty {du\over \exp(-bu)+1}
%% \int \frac{d{\bf q}_t}{(2 \pi)^{d-1}}  e^{-i {\bf q}_t \cdot ({\bf a}_t - {\bf b}_t)  } {\rm Ai}\left(
%% a_n +{\bf q}_t^2 + u \right){\rm Ai}\left(
%% b_n +{\bf q}_t^2 + u \right),  \label{kernelTVedge} \\
%% & =&  
% \int \frac{d{\bf q}_t}{(2 \pi)^{d-1}}  e^{-i {\bf q}_t \cdot ({\bf a}_t - {\bf b}_t)  }  
%K_{\rm Ai}(a_n +{\bf q}_t^2,b_n +{\bf q}_t^2)\;, \label{edgeKd2} 
% \eea
in terms of the Airy kernel $K_{\rm Ai}(a,b) = \int_0^{+\infty} du \, {\rm Ai}(a+u) {\rm Ai}(b+u)$ and the integral over the transverse Fourier variable ${\bf q}_t$ is over the $(d-1)$-dimensional space.
Using the scaling form in Eq. (\ref{kernelTVedge0_app}) into Eq. (\ref{eq_app1}) and writing $a'_n = w_N a_n$ and ${\bf b}' = w_N {\bf b}$ now yields
\begin{equation}
{\rm Tr}( [I_wK]^2) \simeq  S_d \frac{r^{d-1}_{\rm edge}}{w_N^{d-1}}\int_{\tilde w}^\infty da_n \int_{\tilde w}^\infty db_n\int d{\bf b}_t {\cal{K}}_{d,{\bf n}}^{\rm edge}(a_n{\bf n}, b_n{\bf n} + {\bf b}_t)^2,
\end{equation}
where $\tilde w$ is as defined in Eq. (\ref{scaled_w}).

We now use the representation Eq. (\ref{edgeKd2}) twice to represent ${{\cal K}_{d,{\bf n}}^{\rm edge}}^2$, with two transverse Fourier variables ${\bf q}$ and ${\bf q}'$ for each factor. The integral over ${\bf b}_t$ can be carried out yielding a term
$(2\pi)^{d-1}\delta({\bf q}_t+{\bf q}'_t)$ and leads to (see also formula \eqref{product1})
\begin{equation}
{\rm Tr}( [I_wK_\mu]^2) =  S_d \frac{r^{d-1}_{\rm edge}}{w_N^{d-1}(2\pi)^{d-1}}\int_{\tilde w}^\infty da_n \int_{\tilde w}^\infty db_n\int d{\bf q}_t K^2_{\rm Ai}(a_n + q_t^2, b_n + q_t^2).
\end{equation}
The remaining integral over ${\bf q}_t$ can now be carried out and the result written as
\begin{equation}
{\rm Tr}( [I_wK_\mu]^2) =  \frac{S_d S_{d-1}}{2 (2\pi)^{d-1}} (\frac{r_{\rm edge}}{w_N})^{d-1} \int_0^\infty du \,u^{\frac{d-3}{2}}\int_{\tilde w+u}^\infty da_n \int_{\tilde w+u}^\infty db_n K^2_{\rm Ai}(a_n, b_n).
\end{equation}
Using the notation of functional traces we see the the above can be simply written as
\begin{equation}
{\rm Tr}( [I_wK_\mu]^2) =  %\frac{\sqrt{\pi}r^{d-1}_{\rm edge}(d-1)}{(2w_N)^{d-1}\Gamma^2(\frac{d}{2})}
{ \frac{1}{\Gamma(d-1)}  \left(\frac{r_{\rm edge}}{w_N}\right)^{d-1} }
\int_0^\infty du \, u^{\frac{d-3}{2}}{\rm Tr}\left([P_{\tilde w +u}K_{\rm Ai}]^2\right) \;,\label{t2}
\end{equation}
where $P_s$ is the projector on $[s,+\infty[$. 
The computation of higher order traces can be done in an identical manner and it is easy to see that
\begin{equation}
{\rm Tr}( [I_wK_\mu]^n) =  %\frac{\sqrt{\pi}r^{d-1}_{\rm edge}(d-1)}{(2w_N)^{d-1}\Gamma^2(\frac{d}{2})}
{ \frac{1}{\Gamma(d-1)}  (\frac{r_{\rm edge}}{w_N})^{d-1} }
\int_0^\infty du \, u^{\frac{d-3}{2}}{\rm Tr}\left([P_{\tilde w +u}K_{\rm Ai}]^n\right),\label{intu}
\end{equation}
for $n\geq 2$.

In order to analyze the large $\tilde w$ behavior, we use the ``Christoffel-Darboux formula'' for the Airy kernel, in terms of the Airy function ${\rm Ai}$ and its derivative ${\rm Ai}'$,
\begin{equation}
K_{\rm Ai}(x,y) =\frac{ {\rm Ai}(x){\rm Ai}'(y) -{\rm Ai}(y){\rm Ai}'(x)}{x-y},
\end{equation}
and for large $x$ we use the asymptotic formula
\begin{equation}
{\rm Ai}(x) \sim \frac{\exp(-\frac{2}{3}x^{\frac{3}{2}})}{2\sqrt{\pi}x^{\frac{1}{4}}},
\end{equation}
which means that 
\begin{equation}
{\rm Ai}'(x) \sim -\sqrt{x}{\rm Ai}(x) \;.
\end{equation}
Thus, for both $x$ and $y$ large, we have
\begin{equation}
K_{\rm Ai}(x,y) \sim \frac{ {\rm Ai}(x){\rm Ai}(y)}{\sqrt{x}+\sqrt{y}}.\label{asairy}
\end{equation}
Now consider the second order contribution to the trace formula, specifically the last term of the integrand in Eq.~(\ref{t2}) which can be written as 
\begin{equation}
{\rm Tr}\left([P_{\tilde w +u}K_{\rm Ai}]^2\right) = \int_0^\infty da\int_0^\infty db K_{\rm Ai}^2(a+\tilde w + u, b+ \tilde w + u) \;.
\end{equation}
In the above, $\tilde w$ is assumed to be large and $u$ is positive so we can use the asymptotic result from Eq.~(\ref{asairy}) to obtain
\begin{equation}
{\rm Tr}\left([P_{\tilde w +u}K_{\rm Ai}]^2\right) \simeq \int_0^\infty da\int_0^\infty db \frac{{\rm Ai}^2(a+\tilde w + u){\rm  Ai}^2(b+ \tilde w + u)}{(\sqrt{a + \tilde w + u}+ \sqrt{b + \tilde w + u})^2}.
\end{equation}
The asymptotic expansion of the integrand above for large $\tilde w$ is dominated by the exponential term in the Airy function and we obtain
\begin{equation}
{\rm Tr}\left([P_{\tilde w +u}K_{\rm Ai}]^2\right) \simeq \frac{\exp(-\frac{8}{3}\tilde w^{\frac{3}{2}}- 4\tilde w^{\frac{1}{2}}u)}{
64 \pi \tilde w^{\frac{5}{2}}},
\end{equation}
which yields
\begin{equation}
{\rm Tr}( [I_wK_\mu]^2) \sim 
 \left(\frac{r_{\rm edge}}{w_N}\right)^{d-1} 
  \frac{\exp(-\frac{8}{3}\tilde w^{\frac{3}{2}})}{ \tilde w^{\frac{9+d}{4}}}.
% \frac{(d-1)\Gamma(\frac{d-1}{2})\exp(-\frac{8}{3}\tilde w^{\frac{3}{2}})}{2^{2d+4}\sqrt{\pi}\Gamma^2(\frac{d}{2}) \tilde w^{\frac{9+d}{4}}}.
\end{equation}
In the above if we substitute $\tilde w \simeq  a_\mu = \left(\frac{d-1}{2} \ln \mu \right)^{\frac{2}{3}} $ we find that, up to logarithmic corrections
\begin{equation}
{\rm Tr}( [I_wK_\mu]^2) \sim \mu^{-\frac{2}{3}(d-1)}.
\end{equation}
It can be shown by a simple extension of the above argument that
\begin{equation}
{\rm Tr}( [I_wK_\mu]^n) \sim \mu^{-\frac{2(n-1)}{3}(d-1)}
\end{equation}
in the regime of typical fluctuations. Hence, for $d>1$ the higher order traces are irrelevant in that regime.

 \end{document}